\documentclass[epj]{svjour}
\usepackage{latexsym}
\usepackage{color}
\usepackage{graphicx}
\usepackage{appendix}
\usepackage{rotating}
\usepackage{amsmath}

\begin{document}
\title{Systematics of toroidal dipole modes in Ca, Ni, Zr, and Sn isotopes}
\author{A. Repko\inst{1},  V.O. Nesterenko\inst{2,3,4}, J. Kvasil\inst{5},
\and P.-G. Reinhard\inst{6}}
\institute
{Institute of Physics, Slovak Academy of Sciences, 84511 Bratislava, Slovakia
\and
Laboratory of Theoretical Physics,
Joint Institute for Nuclear Research, Dubna, Moscow Region, 141980, Russia
\and
State University "Dubna", Dubna, Moscow Region, 141980, Russia
\and
Moscow Institute of Physics and Technology,
Dolgoprudny, Moscow region, 141701, Russia
\and
Institute of Particle and Nuclear Physics, Charles University,
CZ-18000 Prague 8, Czech Republic
\and
Institut f\"ur Theoretische Physik II, Universit\"at Erlangen, D-91058, Erlangen, Germany
}
\date{Received: date / Revised version: date}

\abstract{We analyze the relation between isoscalar toroidal modes
  and so-called pygmy dipole resonance (PDR), which both appear
  in the same region of low-energy dipole excitations. To this end, we
  use a theoretical description within the fully self-consistent
  Skyrme quasiparticle random-phase approximation (QRPA). Test cases
  are spherical nuclei $^{40,48}$Ca, $^{58,72}$Ni,
  $^{90,100}$Zr, and $^{100,120,132}$Sn which cover four different
  elements and for each element at least two isotopes with different
  neutron excess, one small and another large. The structure of
  the modes is investigated in terms of strength functions,
  transition densities (TD) and transition currents (TC).  For all considered
  nuclei, we see that, independently on whether PDR strength exists or not,
  the flow pattern in the lower part of the "PDR
  energy region" is basically an isoscalar vortical toroidal motion
  with a minor irrotational fraction. A one-to-one correspondence between
  calculated TD and TC is established. The toroidal flow appears
  already in the uncoupled two-quasiparticle (2qp) excitations and
  becomes definitively strong for the QRPA modes.
 Altogether, we find that low-lying dipole strength often denoted as isoscalar
 PDR is actually an oversimplified imitation of the basically toroidal
 motion in nuclei with sufficient neutron excess.}

\maketitle

\section{Introduction}

The dipole toroidal mode in nuclei represents a remarkable example of
confined vortical flow \cite{Se81,Pa07,Ne16PAN}. This
mode forms a vortical ring, called a Hill’s vortex in hydrodynamics
\cite{Hill1894,MiTho60}.  In 1983, S.F. Semenko has predicted the existence of
a toroidal dipole resonance (TDR) in nuclei located at an energy $E
\approx (50-70) A^{-1/3}$ MeV \cite{Se81}. During subsequent decades,
TDR was a subject of intense theoretical studies both in macroscopic and
self-consistent microscopic models, see e.g.
\cite{Bas93,Balb94,Mis06,Co00,Vr02,Rich04,Papa11} and more
references in the reviews
\cite{Pa07,Ne16PAN}. There are numerous experimental data from
isoscalar ($\alpha,\alpha’$) scattering
\cite{Morsch_80,Adams_86,Davis_97,Clark_01,Youngblood_04,Uchida_PL_03,Uchida_PRC_04}
and nuclear fluorescence \cite{Ryezayeva_02} which claim that TDR forms
the low-energy part of the isoscalar giant dipole resonance (IS GDR)
while the high-energy part of IS GDR is ascribed to the irrotational
compression dipole resonance (CDR), for reviews see \cite{Pa07,Har01}.
Schematic vortical TDR and irrotational CDR velocity fields are shown in Fig. 1.
However, despite impressive theoretical and experimental efforts,
some important features of the TDR are still under debate and its
unambiguous experimental observation has yet to come, see e.g. the
discussion in \cite{Rep17EPJA}.
\begin{figure}
\begin{center}
  \includegraphics[width=0.9\columnwidth,angle=-0]{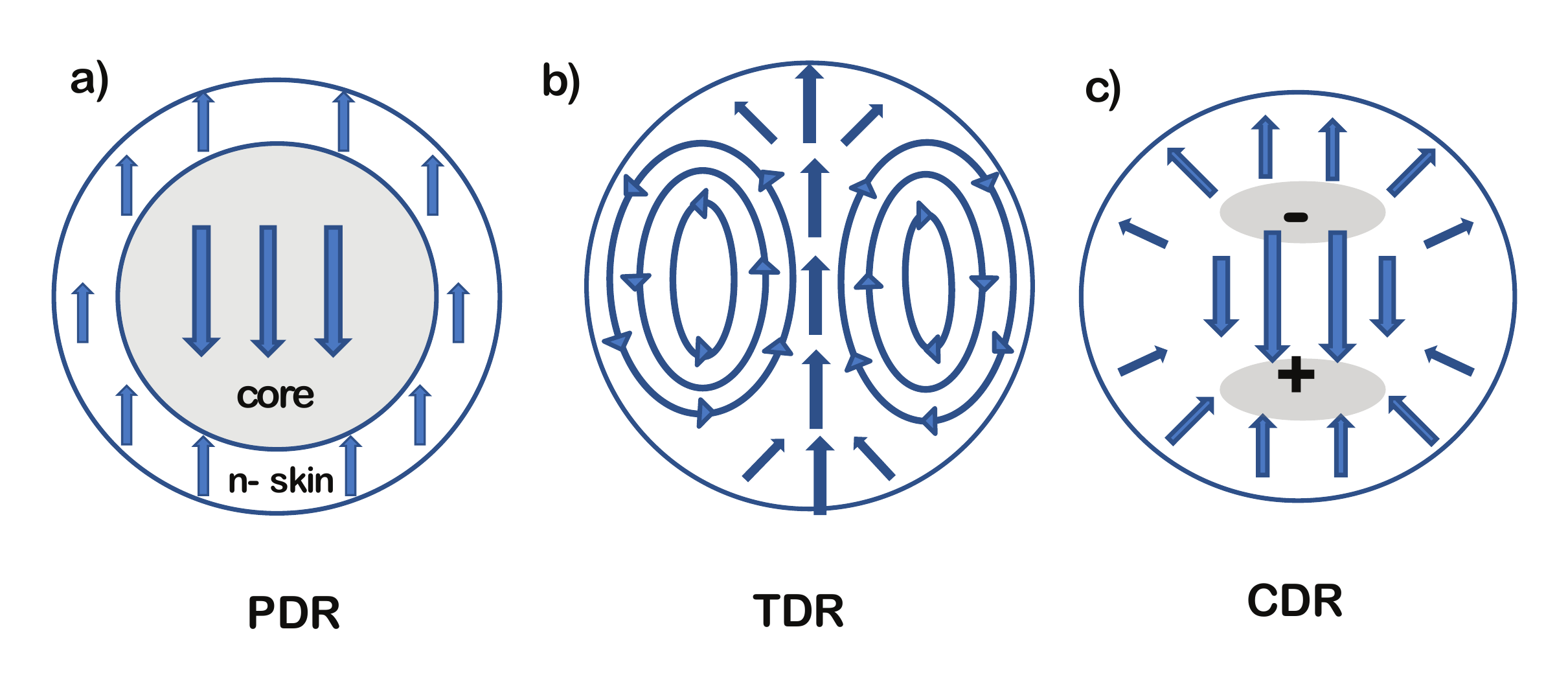}
\vspace{0.6cm}
\caption{Sketch of velocity fields for PDR (a), TDR
  (b) and CDR (c) flows.  The arrows indicate direction (not strength)
  of the flows. In the plot (c), compression (+) and decompression (-)
  regions with increased and decreased densities are marked.
  }
\end{center}
\label{fig:schem}       
\end{figure}

During last decade, our collaboration studied continuously various
features of the vortical dipole toroidal mode
\cite{Ne16PAN,Rep17EPJA,Kv11,Rep13,Ne15Dre,Rei14vor,Kv13Sm,Kv14Yb,Ne18PRL,Ne18EPJWC,RepSEARPA}
using separable \cite{Ne02,Ne06} and non-separable \cite{Rep17EPJA,Repcode} versions
of the fully self-consistent Quasiparticle Random Phase Approximation
(QRPA) approach with Skyrme forces. The key
observables of the TDR, namely strength functions, transition
densities (TD), and transition current (TC),
were analyzed in spherical and deformed nuclei.
TDR and CDR were analyzed with respect to the familiar criteria of nuclear
vorticity \cite{Rei14vor}.
A new route in exploration of the vortical
toroidal mode in nuclei was recently proposed \cite{Ne18PRL,Ne18EPJWC}. It was shown that
there should exist low-lying individual toroidal dipole states well separated
from neighboring excitations in light axially-deformed nuclei.
These states can be easier observed and identified than in the
swamp of low-energy modes in heavier nuclei.
Similar results for light nuclei were obtained within the approaches
involving cluster degrees of freedom
\cite{KE10Be_PRC17,KE12C_PRC18,KE_rew18,KE10Be_arXiv,KE16O_arXiv}.

The present study will concentrate in one of the most interesting and
significant aspects concerning the toroidal mode: the relation between
isoscalar TDR and low-energy part of so  called pygmy
dipole resonance (PDR). It is known that PDR is split into the low-lying
isoscalar (IS) fraction, often interpreted by neutron-skin oscillations, and a
higher isovector (IV) part \cite{End10,Sa13,Br19}. The TDR also has IS and IV
components \cite{Kv11}.  As shown in our previous works
\cite{Ne16PAN,Rep13,Ne15Dre}, the IS PDR can be viewed as a local
manifestation of the IS toroidal mode on the nuclear surface showing up in
nuclei with sufficient neutron excess $(N > Z)$. Indeed, IS TDR and PDR
lie in the same energy region and our calculated current distributions
for dipole states in the IS PDR
region show clear toroidal pattern \cite{Ne16PAN,Rep13,Ne15Dre}
schematically illustrated in plot (b) of Fig. 1. Besides, the analysis
of strength functions and TD  indicates  that
dipole states in the IS PDR region also have a minor compressional fraction
\cite{Kv11,Rep13,Reinhard_2013}.
While the TC is dominated by vortical toroidal flow, the TD is perhaps
mainly  determined by the irrotational compressional fraction.
For the latter, see also results of the correlation analysis
\cite{Reinhard_2013}.

Our previous analysis of the relation between IS TDR and PDR was
performed only for a few selected nuclei: $^{120}$Sn \cite{Ne16PAN},
$^{208}$Pb \cite{Rep13,Ne15Dre}, and $^{154}$Sm \cite{Rep17EPJA}.  Here we
aim at a more systematic survey and extend our exploration to a wider
set of nuclei, namely to Ca, Ni, Zr and Sn isotopes ($40 \le A \le 132$). For
each element, isotopes with and without neutron excess will be
compared.  Our goal is to show that low-energy dipole spectra in all
considered nuclei, independently of their neutron excess, represent
predominantly the toroidal current distribution with some irrotational fractions
and that, in nuclei with $N > Z$, this flow appears also as
PDR peaks.  We will demonstrate that the neutron flow at the nuclear
boundary can be roughly imitated by a PDR-like irrotational
pattern, see panel (a) in Fig. 1. Note that our study addresses only IS
parts of TDR and PDR. Following \cite{Kv11}, the IV TDR pattern
is more complicated and its comparison with IV PDR requires a separate analysis.

The paper is organized as follows: In Section 2, the theoretical and calculation
background is sketched. In Sec. 3, results of the calculations for Ca, Ni, Zr,
and Sn isotopes are discussed. The low-energy dipole spectrum in $^{120}$Sn is
scrutinized in most detail.
In Sec. 4, the features of TDR and PDR are compared
and possible reactions for observation of the vortical toroidal mode
are briefly discussed. Conclusions are given in Sec. 5. Appendix A
includes definitions of the density and current operators with the relevant
effective charges. In Appendix \ref{sec:120Sn}, the calculated and observed
dipole strengths in $^{120}$Sn are compared.
A list of acronyms is provided in Appendix \ref{sec:acro}.

\section{Theoretical and calculation background}
\label{sec:theor}

The calculations are performed within
QRPA
\cite{Ri80} based on the Skyrme energy functional \cite{Vau,Be03}
\begin{equation}
\mathcal{E}(\rho,\tau, {\bf J}, \vec{j}, \vec{s}, \vec{T}, \tilde{\rho})
= \mathcal{E}_{\rm kin} + \mathcal{E}_{\rm Sk}
+ \mathcal{E}_{\rm Coul} + \mathcal{E}_{\rm pair}
\label{Efunc}
\end{equation}
including kinetic, Skyrme, Coulomb and pairing parts.
The Skyrme part $\mathcal{E}_{\rm{Sk}}$ depends on the following local
densities and currents: density $\rho(\vec{r})$, kinetic-energy
density $\tau(\vec{r})$, spin-orbit density $\mathbf{J}(\vec{r})$,
current $\vec{j}(\vec{r})$, spin density $\vec{s}(\vec{r})$, and spin
kinetic-energy density $\vec{T}(\vec{r})$. The Coulomb part includes
direct and exchange term, the latter is treated in Slater
approximation.  The pairing functional depending on the pairing density
$\tilde{\rho}(\vec{r})$ is derived from a zero-range
contact interaction \cite{Be00}. Here we use mainly volume pairing
(without  dependence on density $\rho(\vec{r})$) treated at the BCS level
\cite{Rep17EPJA}.

The mean-field Hamiltonian and QRPA residual interaction are
determined through the first and second functional derivatives of the
total energy (\ref{Efunc}), see e.g. \cite{Reinhard_92}.
We do not employ the separable ansatz for the residual
interaction, as done in earlier papers \cite{Kv11,Repcode},
but use full QRPA solved as matrix equations in configuration space.
The approach
is fully self-consistent since: i) both the mean field and residual
interaction are obtained from the same Skyrme functional, ii)
time-even and time-odd densities are used, iii) the residual
interaction includes all the terms of the initial Skyrme functional as
well as the Coulomb direct and exchange terms, iv) both ph- and
pp-channels in the residual interaction are taken into
account.

Most calculations are performed with the Skyrme force SLy6
\cite{SLy6}.  Earlier, this force was successfully used for the
systematic exploration of IV-GDR in rare-earth, actinide and
superheavy nuclei \cite{Kl08}. Since our study is mainly applied to
dipole excitations, we continue to use SLy6. To check the dependence
of our results on Skyrme parametrizations, we counter check the case
with $^{120}$Sn using the forces SV-bas \cite{SVbas} and SkM* \cite{SkMs}.
SLy6 and SkM* calculations are performed with the simple volume
pairing. SV-bas is defined with surface pairing and so that
is used in this particular case.

We employ a large two-quasiparticle (2qp) basis ranging up to 120 MeV
in $^{120}$Sn. With this basis, the Thomas-Reiche-Kuhn sum rule
\cite{Ri80,Ne08} and isoscalar energy-weighted sum rule \cite{Har01}
are exhausted by 99 - 103 \%.

The relevant characteristics for our analysis are strength functions,
transition densities, and current transition densities.
The strength function for $E1$ transitions between QRPA ground state
$| 0 \rangle$ and excited state $|\nu\rangle$ reads
\begin{eqnarray}
\label{13}
&&S_X(E1, T; E) = \sum_{\mu \ge 0 }
 (2-\delta_{\mu,0})
 \\
&\cdot & \sum_{\nu} [E_{\nu}]^{m_X} \: \big| \langle \nu|\:
\hat{M}^X_{1\mu}(T) \: |{0} \rangle \big|^2 \: \xi_{\Delta}(E-E_{\nu})
\nonumber
\end{eqnarray}
where $E$ is the excitation energy, $\hat{M}^X_{1\mu}(T)$
is the transition operator,
$\mu$ is its azimuthal angular momentum,
$X=\{\rm el, com, tor\}$ marks the type
of the excitation (isovector $E1(T=1)$, compressional isoscalar
$E1(T=0)$, toroidal isoscalar $E1(T=0)$), $E_{\nu}$ is
the energy of the QRPA state. We have the energy factor ($m_X$=1) for
$X={\rm el}$ and no factor ($m_X$=0) for $X={\rm com, tor}$.
The effective charges for transition operators $\hat{M}^X_{1\mu}(T)$
are given in Appendix A.

The strength function  is weighted by a Lorentz function
\begin{equation}
\xi_{\Delta}(E-E_{\nu}) = \frac{1}{2 \pi}
\frac{\Delta (E)}{(E-E_{\nu})^2 + [\Delta(E)/2]^2}
\label{17}
\end{equation}
with energy-dependent folding \cite{Kv13Sm}
\begin{equation}
\Delta(E) =
\left\{
\begin{array}{ll}
 \Delta_0 & \:{\rm for} \:\:\: E\leq E_0, \\
 \Delta_0 + a\:  (E-E_0) &\: {\rm for} \:\:\: E> E_0.
\end{array} \right.
\label{18}
\end{equation}
The Lorentzian folding with $\xi_{\Delta}(E-E_{\nu})$ is used
to simulate the escape width and the coupling to the complex configurations
(spreading width). Since these two effects grow with increasing
excitation energy, we employ an energy dependent folding width
$\Delta(E)$ with the parameters $\Delta_0$=0.3 MeV,
$E_0 = {\rm min}\{S_n,S_p\}$ MeV (with $S_n$ and $S_p$ being neutron
and proton separation energies) and $a=$0.1667 \cite{Rep17EPJA}.

The IV dipole operator $(X={\rm el})$ reads
\begin{equation}
\hat{M}^{\rm el}_{1\mu}(T=1) =
 e \sum_{q=n,p} e_{\rm eff}^q
 \sum_{i \in q}
\: r_i  Y_{1\mu}(\Omega_i)
\label{14}
\end{equation}
where $Y_{1\mu}(\Omega_i)$ is the spherical harmonic and
$e_{\rm eff}^q$ are effective charges
equal to $N/A$ for protons (q=p) and $-Z/A$ for neutrons (q=n).
For this operator, the photoabsorption is
\begin{equation}
 \sigma_{\gamma}(E1,T=1;E) = 0.402 \; S_{\rm el}(E1,T=1; E)\; [{\rm fm}^2].
\end{equation}

The toroidal IS dipole operator reads \cite{Kv11}
\begin{eqnarray}
&&
\hat{M}^{\rm tor}_{1\mu}(T=0) =
 - \frac{1}{10c\sqrt{2}}
\int d\vec{r} \:
r^{3} {\bf Y}_{11 \mu}(\Omega_i)\cdot
(\nabla \times \hat{{\bf j}}_{\rm nuc})
\nonumber \\
&&
\qquad \qquad \quad \quad \; \;
= - i \frac{1}{2c\sqrt{3}}
\int d\vec{r} \:
\hat{{\bf j}}_{\rm nuc}
\nonumber \\
&&
\qquad \cdot
\big[
\frac{\sqrt{2}}{5} r^2{\bf Y}_{12\mu}(\Omega)
+ (r^2 - \langle r^2 \rangle_0)
{\bf Y}_{10\mu}(\Omega)
\big]
\label{15}
\end{eqnarray}
where
$\hat{\bf j}_{\rm nuc}({\bf r})=
\hat{\bf j}_{\rm c}({\bf r})+\hat{\bf j}_{\rm m}({\bf r})$
is the operator of  nuclear current including convective and magnetization
parts, see the explicit expression in the Appendix A, and
 ${\bf Y}_{10\mu}(\Omega)$ and ${\bf Y}_{12\mu}(\Omega)$ are vector spherical harmonics.
The terms with the ground-state square radius $\langle r^2 \rangle_0$
take care of the center-of-mass corrections \cite{Har01,Kv11,RepSEARPA}.
The toroidal operator (\ref{15}) is formally a sum of vortical and
irrotational (compressional) parts \cite{Kv11,Rei14vor}. However, the
low-energy response (\ref{13}) from this operator is mainly vortical
 \cite{Kv11,Rei14vor}.

The operator for IS compressional dipole
can be written in terms of the nuclear current \cite{Har01,Kv11},
\begin{eqnarray}
&&
\hat{M}^{\rm com}_{1\mu}(T=0) =
 - i \frac{1}{10c}
\int d\vec{r} \:
r^{3} Y_{1 \mu}(\Omega)
(\nabla \cdot \hat{{\bf j}}_{\rm nuc})
 \nonumber\\
&&
\qquad \qquad \qquad \quad
= - i \frac{1}{2c\sqrt{3}}
\int d\vec{r} \: \hat{{\bf j}}_{\rm nuc}
\nonumber \\
&&
\qquad
\cdot \big[
\frac{2\sqrt{2}}{5} r^2 {\bf Y}_{12\mu}(\Omega)
- (r^2 - \langle r^2 \rangle_0)
{\bf Y}_{10\mu}(\Omega)
\big] \; ,
\label{15-CM}
\end{eqnarray}
or, using the continuity equation, in the familiar form
\begin{equation}
\label{Tcom}
\hat{M'}^{\rm com}_{1\mu} (T=0)
= \frac{e}{20} \:
\sum_{i=1}^{A} [ r_i^3
-\frac{5}{3}\langle r^2 \rangle_0 r_i ]
Y_{1 \mu}(\Omega_i) .
\end{equation}
The compressional operator is irrotational \cite{Kv11,Rei14vor}.

In spherical nuclei, the transition densities (TD) and transition currents (TC) for
the transfer from the ground state $I^{\pi}\mu=0^+0$ to the excited
state  $\nu$ with $1^-\mu$ read \cite{RW87}
\begin{eqnarray}
\label{TD}
\delta \rho_{\nu\mu}({\bf r}) &=& \langle\nu|\hat{\rho}({\bf r})|0\rangle
= \delta\rho_{\nu}(r) Y^*_{1\mu}(\Omega)
\\
\label{TC}
\delta {\bf j}_{\nu\mu}({\bf r}) &=& \langle\nu|\hat{{\bf j}}'_c({\bf r})|0\rangle
= i  \sum_{L=0,2} j^{1L}_{\nu}(r){\bf Y}^*_{1L\mu}(\Omega)
\end{eqnarray}
where $\hat{\rho}({\bf r})$ is the density operator
and $\hat{\bf j}'_{\rm c}({\bf r})=m/(e \hbar) \hat{\bf j}_{\rm c}({\bf r})
$ is the scaled operator
of the convection nuclear current, see definitions in the Appendix A.
For simplicity, we use in TC plots only the convection current.
The TD and irrotational TC are related by the continuity equation
\begin{equation}\label{CE}
    - i\frac{m}{\hbar^2}E_{\nu} \delta{\rho}_{\nu\mu}=\nabla \cdot \delta{\bf j}_{\nu\mu} .
\end{equation}
where
\begin{equation}
\nabla \cdot \delta{\bf j}_{\nu\mu} =i [\mathrm{div}\;\delta j]_{\nu}(r) Y^*_{1\mu}(\Omega)
\end{equation}
and
\begin{equation}\label{rad_div}
[\mathrm{div}\; \delta j]_{\nu}(r)=\sqrt{\frac{1}{3}}\frac{d}{dr} j^{10}_{\nu}(r) -
 \sqrt{\frac{2}{3}}[\frac{d}{dr} +\frac{3}{r}]j^{12}_{\nu}(r) .
\end{equation}
\begin{figure*} 
\begin{center}
  \includegraphics[width=1.9\columnwidth,angle=-0]{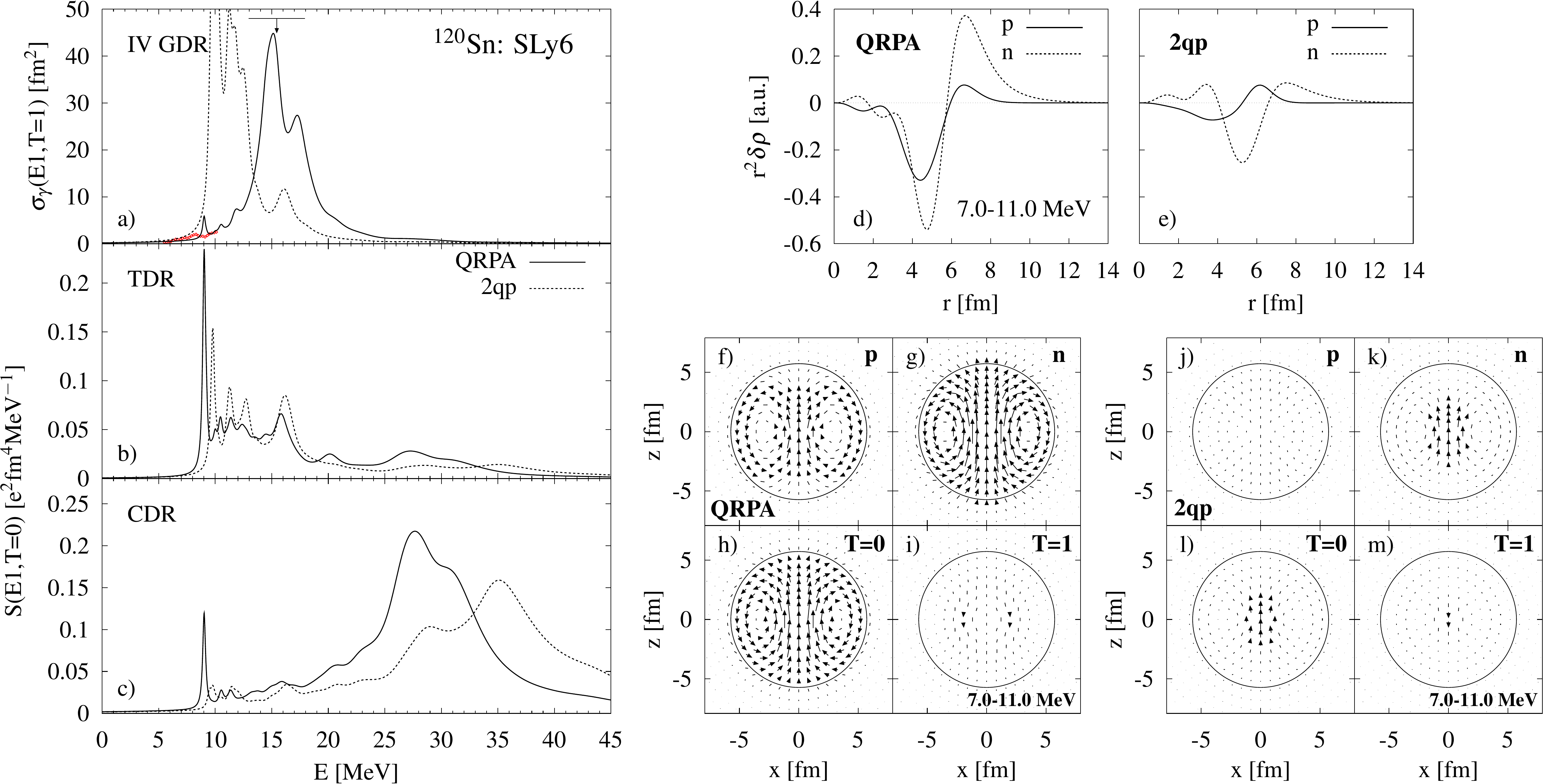}
\caption{(Color online) Results for $^{120}$Sn computed with SLy6. Upper left block:
  IV dipole, IS toroidal, and IS compressional strength
  distributions. The experimental dipole strength at E=5.5--10 MeV
  \cite{Kru15_120Sn} is marked by red open rhombuses. The experimental
  data \cite{Lep74} for IV GDR, maximum-strength energy and FWHM,
  are depicted by vertical arrow and horizontal line, respectively.
  Upper right block: TD integrated over the energy interval 7--11 MeV,
  left panel from QRPA  and right panel from mere 2qp states. Lower right
  block: TC for proton, neutron, isoscalar ($T$=0), and
  isovector ($T$=1) flows; the flow pattern are shown for QRPA and mere
  2qp states as indicated.
}
\end{center}
\label{fig:120SnSLy6}       
\end{figure*}

The radial TD $\delta\rho_{\nu}(r)$ and divergence of TC
$[\mathrm{div}\; \delta j]_{\nu}(r)$ as well as  angular-dependent TC
(\ref{TC}) for $\mu=0$  will be plotted and analyzed  in
Sec. 3 (in what follows we will omit the index $\mu$).
To suppress individual details of the states and
highlight their general features, these variables are averaged over
the dipole states in the chosen  energy interval $[E_1, E_2]$
following the prescription \cite{Rep13,Rei14vor}.
The averaging is done by summing the variables weighted by the
dipole matrix elements
$D_{\nu}=\langle\nu|\hat{M}^{\rm com}_{10} (T=0)|0\rangle$.
This amounts to
\begin{subequations}
\label{eq:averDC}
\begin{eqnarray}
\label{sum_td}
 \delta\rho(r)
 &=&
 \sum_{\nu\in[E_1, E_2]}
  D_{\nu}^* \; \delta \rho_{\nu}(r)
  \;,
\\
\label{sum_tcd}
  \delta {\bf j}({\bf r})
  &=&
   \sum_{\nu\in[E_1, E_2]}
  D_{\nu}^* \; \delta {\bf j}_{\nu 0}({\bf r}) ,
\\
 \label{sum_div}
 [\mathrm{div}\; \delta j](r)
  &=&
   \sum_{\nu\in[E_1, E_2]}
  D_{\nu}^* \; [\mathrm{div}\; \delta j]_{\nu}(r) ,
\end{eqnarray}
\end{subequations}
where $\delta \rho_{\nu}(r)$,
$\delta {\bf j}_{\nu 0}({\bf r}) $ and
$[\mathrm{div}\; \delta j]_{\nu}(r)$
are given by Eqs.  (\ref{TD}), (\ref{TC}) and (\ref{rad_div}) using the sets of
the effective charges corresponding to the proton, neutron, IS and IV transitions.
The sets are listed  in the Appendix A.
The $D_{\nu}^*$ factors associate the proper weight to the contributions and render the
expressions bilinear in $|\nu\rangle$ which, in turn, removes nonphysical
dependence on the phase of $|\nu\rangle $, see \cite{Rep13,Rei14vor} for more detail.
In the present study, the averaged TD and TC are calculated for both
QRPA and 2qp states at the given energy interval.

In TD and TC from (\ref{TD})-(\ref{TC}),
the spurious admixtures arising due to the motion of the nuclear center-of-mass
are eliminated following the prescription \cite{RepSEARPA}.
The unperturbed 2qp cases are corrected separately for protons and neutrons.

\section{Results of the calculations}

As a first example from our large selection of nuclei, we look at the
case of $^{120}$Sn computed with SLy6. This case is shown in Fig. \ref{fig:120SnSLy6}.
Its left part collects the strength functions for IV GDR, IS TDR and IS CDR
(panels (a)-(c)). The compression strength is calculated with
the current-dependent operator (\ref{15-CM}).
In  the right upper panels (d)-(e), we show the averaged
proton and neutron radial TD  $\delta\rho (r)$ from Eq. (\ref{sum_td}).
In the lower right panels (f)-(m), we demonstrate the averaged TC flow pattern
from Eq. (\ref{sum_tcd}). Both TD and TC are averaged over the
PDR region 7-11 MeV. For each observable, the
results from QRPA and unperturbed 2qp states are compared.

Following the panel (a), we well describe the experimental data
\cite{Lep74} for the peak energy and FWHM of the IV GDR,
which confirms the accuracy of our model.
Further, panels (a)-(c)  show that the PDR energy region coincides with
the location of the low-energy parts of the toroidal and compressional
strengths.  This indicates that the nature of dipole states in this
region is complicated.  They should have both large vortical (TDR) and
minor irrotational (probed by the compression dipole external field)
fractions. In general, the IS TDR covers a large energy region 8-18 MeV.
We will analyze only the low-energy part of the region, where both IS
modes of our interest, TDR and PDR, coexist.

The residual interaction of QRPA downshifts
the lowest toroidal and compressional peaks and considerably enhances
their strength (and thus collectivity). At the same time, following
our analysis, these dipole states still maintain rather large 2qp
components (see Appendix B for more detailed information on the calculated
dipole states in the PDR energy region and comparison of the obtained
E1 strength with the experimental data \cite{Kru15_120Sn}).
The strong effect of the residual interaction is also seen in panels (d)-(e)
for the radial TD: the neutron TD dominates at the nuclear surface for
QRPA (which is often considered as justification of the a
collective flow associated with a ``pygmy resonance'') but not for 2qp case.

Complementing and more detailed information on the nuclear flow is
provided by the angular-dependent TC shown in panels (f)-(m). In the 2qp
case, neutron and $T$=0 pattern show weak signs of toroidal flow which then
becomes very pronounced in QRPA.  The dominant collective flow in
the interval 7-11 MeV is obviously IS toroidal, see for comparison
Fig. 1(b). The main contribution comes from neutrons in  accordance with the dominance
of neutron TD at the nuclear surface. We may state that the schematic
picture of PDR as neutron oscillations against an $N=Z$ core is actually
a crude map of the true mechanism behind, toroidal flow of
neutrons at the nuclear boundary.

To gain a deeper insight into the relation between TDR and PDR, we
present in Figs. 3-4 the calculated TD and TC in more detail.
The main question is: can one explain the behavior of the neutron TD,
in particular its maximum at the nuclear surface, by the neutron current
which is basically toroidal?

\begin{figure} 
\begin{center}
  \includegraphics[width=1\columnwidth,angle=-0]{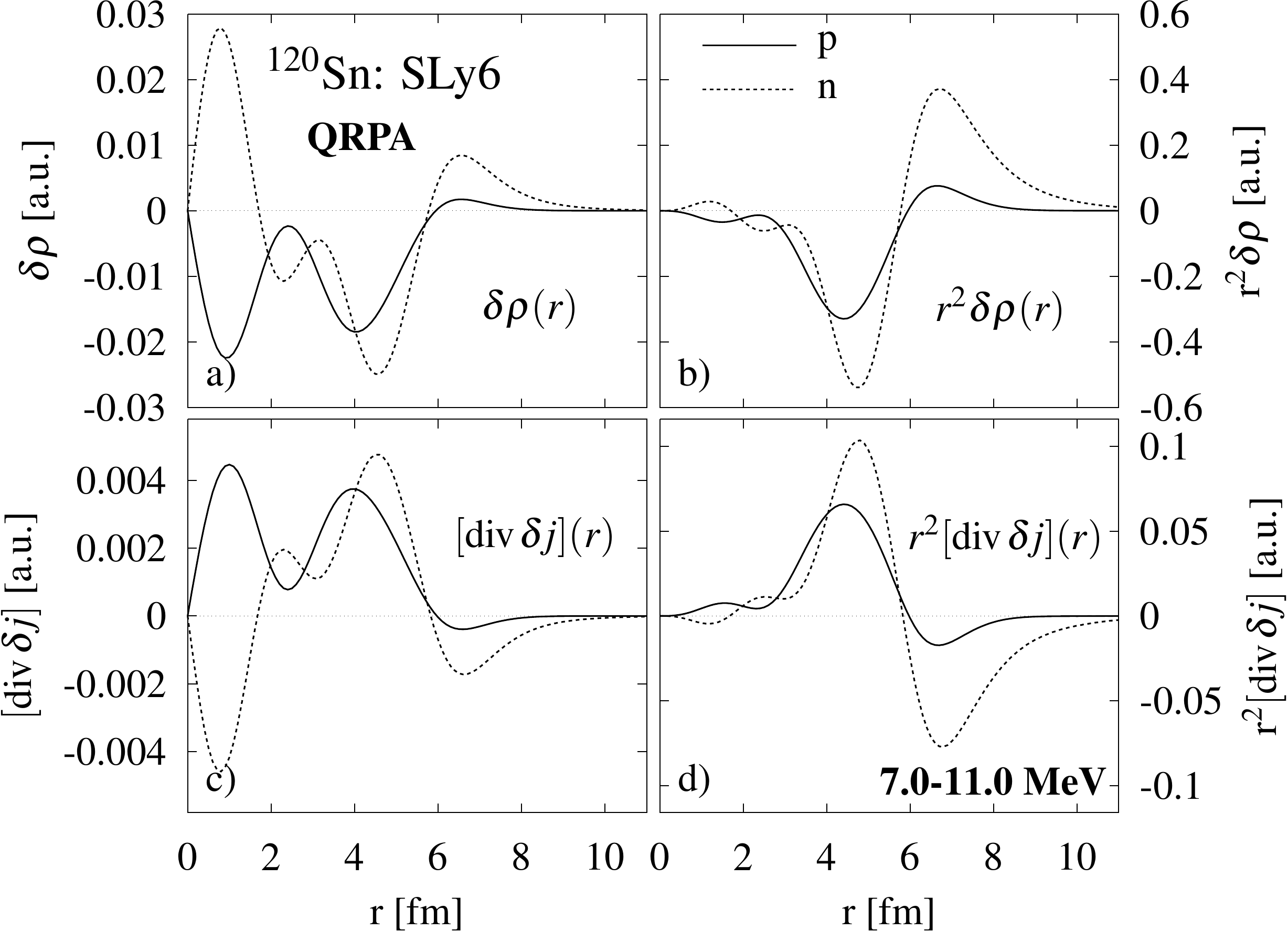}
\caption{Upper plots: averaged radial proton and neutron TD for
QRPA (SLy6) states at the region 7--11 MeV in $^{120}$Sn.
Left and right panels exhibit TD without and with $r^2$-weight.
Bottom plots: the same but for the averaged radial part of the divergence
of the nuclear current.}
\end{center}
\label{fig:3}       
\end{figure}

In the upper panels of Fig. \ref{fig:3}, we show the neutron
and proton  TD with and without $r^2$-weight, which
allows to discriminate the TD behavior at the nuclear surface and
deep inside. We see that the neutron TD without $r^2$-weight (panel (a))
exhibits three humps: positive at 0.5 -- 2 fm, negative at 4 -- 6 fm,
and positive at 6 -- 8 fm.  In neutron TD with $r^2$-weight (panel (b)), the
first hump is much suppressed and the third hump at the surface is
enhanced. Obviously, these tree TD humps should be related to some
essential structures in TC.

We know that TD and TC are linked by the continuity equation (CE)
(\ref{CE}). The nuclear current enters the CE through its divergence. So only
the {\it irrotational} fraction of TC contributes to CE while the dominant
{\it vortical} toroidal  current
${\bf j}_{\rm tor} \sim  \nabla \times \big(\nabla
\times ({\bf r}{\rm M}^{\rm com}_{1\mu}({\bf r}))\big)$
\cite{Bas93,Mis06,Kv11} does not. The CE holds
for each QRPA $\nu$-state. However, the CE
may be somewhat distorted for the {\it averaged} TD
and TC (\ref{eq:averDC}) which are just the variables of our interest.
To check this point, we exhibit in the lower panels of
Fig. \ref{fig:3}  the radial parts $[\mathrm{div}\; \delta j](r)$ of the divergence
of the averaged proton and neutron currents. We see that the behavior of the
divergences strictly correlates with the behavior of the TD.
So CE basically holds and the humps in the TD represent the
inherent features of the actual (basically toroidal) TC.
\begin{figure} 
\begin{center}
  \includegraphics[width=0.9\columnwidth,angle=-0]{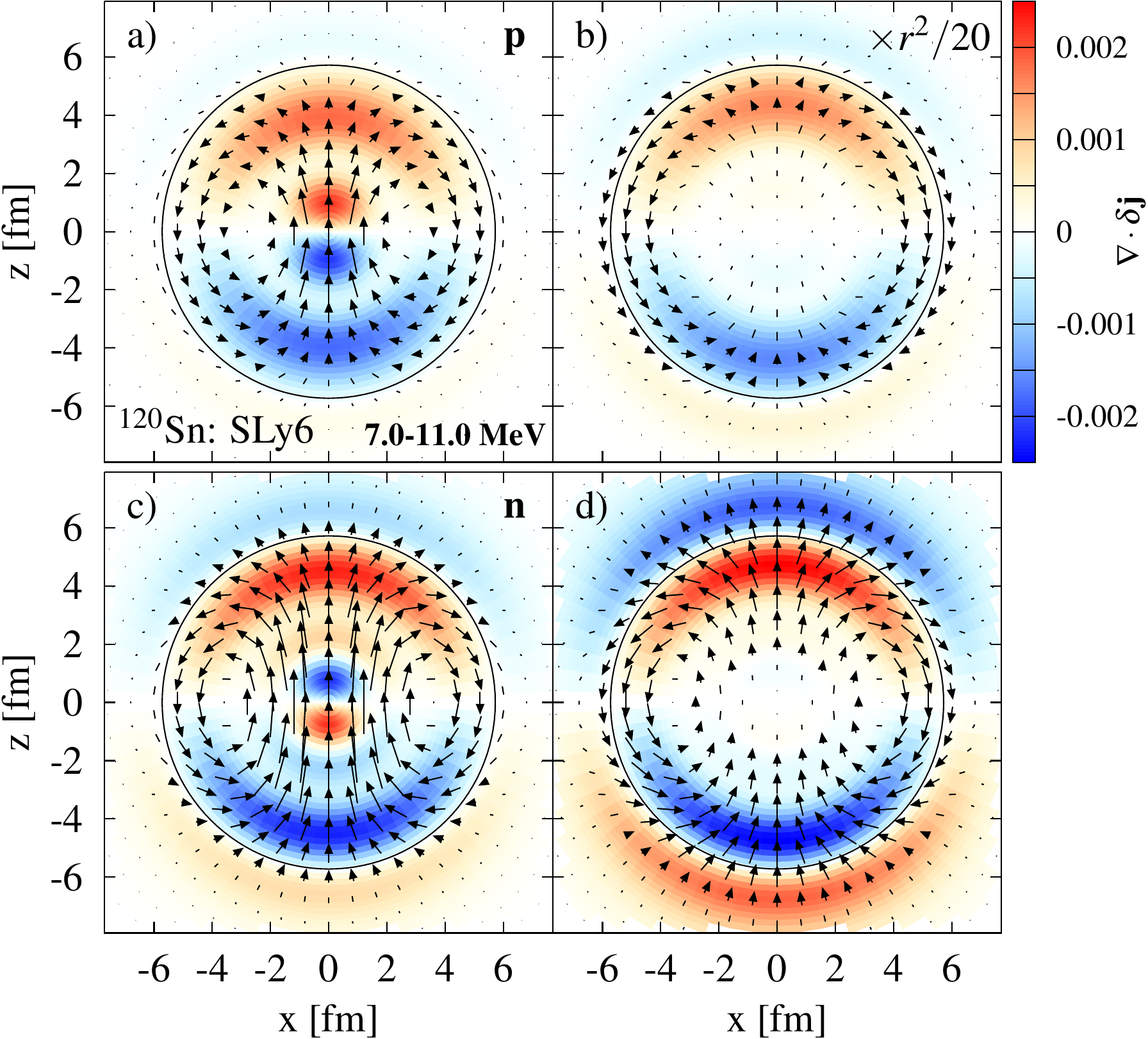}
\caption{(Color online) The averaged proton (upper panels) and neutron (bottom panels)
TC and their angular-dependent divergences for QRPA (SLy6) states at the region
7--11 MeV. The variables are exhibited in
(x, y=0, z)-plane. Like in Fig. 2, the TC are depicted by arrows.
The divergences are exhibited by color (tone) spots  as indicated.
Left and right panels show the variables without and with $r^2$-weight.
For a better view, the arrows for TC and color scaling with $r^2$-weight in the right panels are
decreased by the factor 1/20.
}
\end{center}
\label{fig:4}       
\end{figure}
\begin{figure*} 
\begin{center}
  \includegraphics[width=1.8\columnwidth,angle=-0]{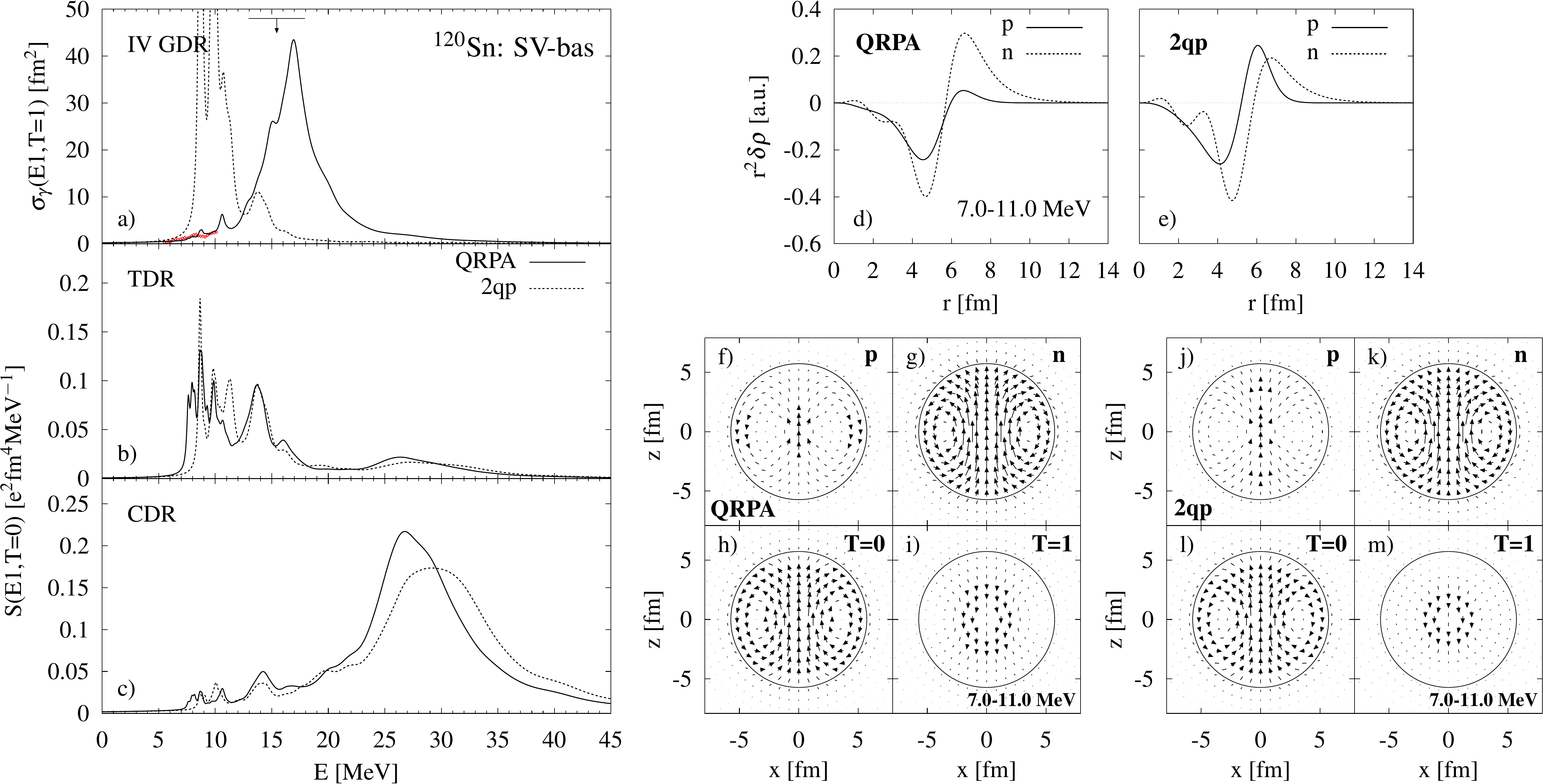}
\caption{(Color online) As in figure 2, but computed with SV-bas.}
\end{center}
\label{fig:5}       
\end{figure*}
\begin{figure*} 
\begin{center}
  \includegraphics[width=1.8\columnwidth,angle=-0]{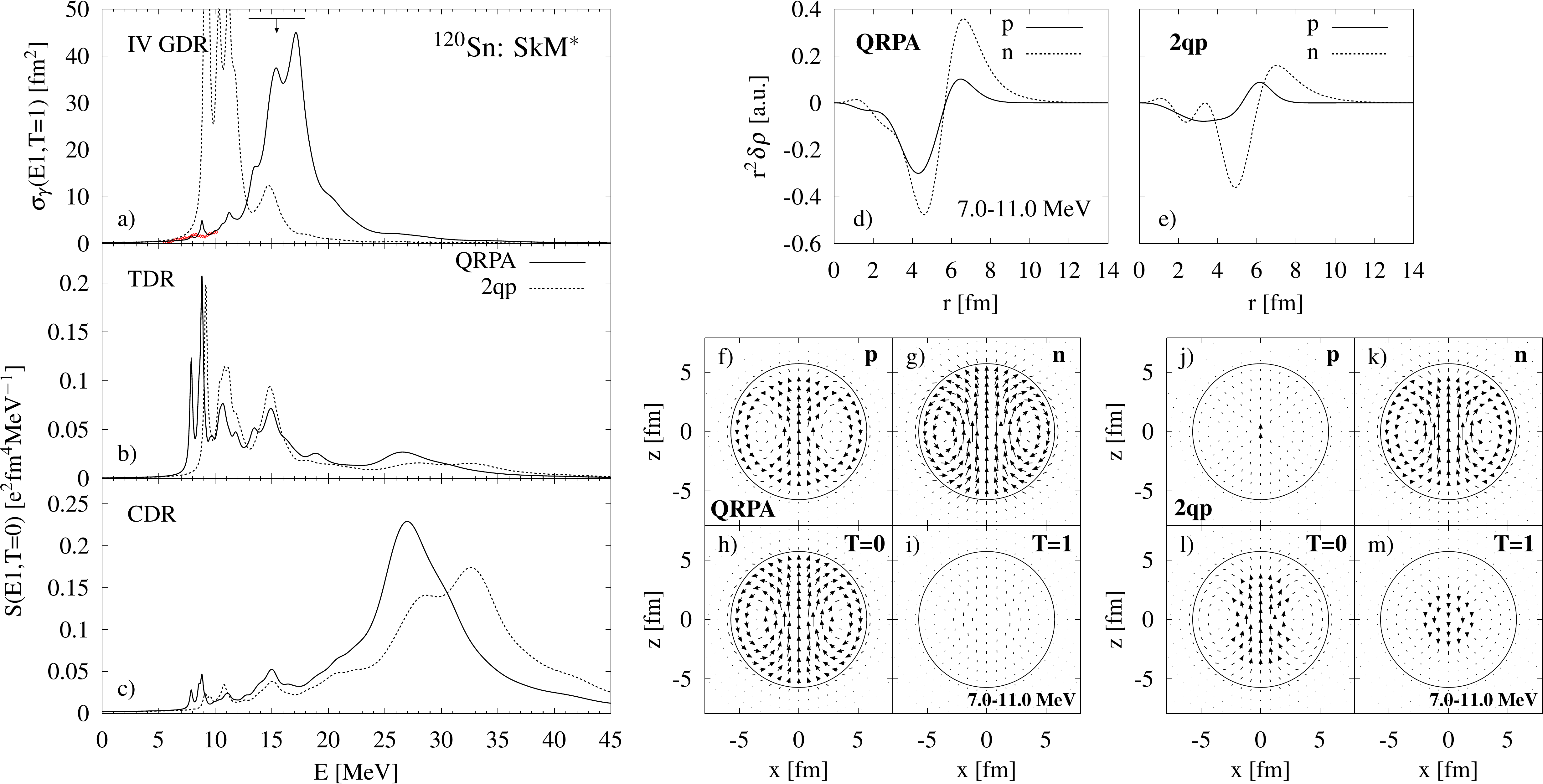}
\caption{(Color online) As in figure 2, but computed with SkM*.}
\end{center}
\label{fig:6}       
\end{figure*}
\begin{figure*} 
\begin{center}
  \includegraphics[width=1.8\columnwidth,angle=-0]{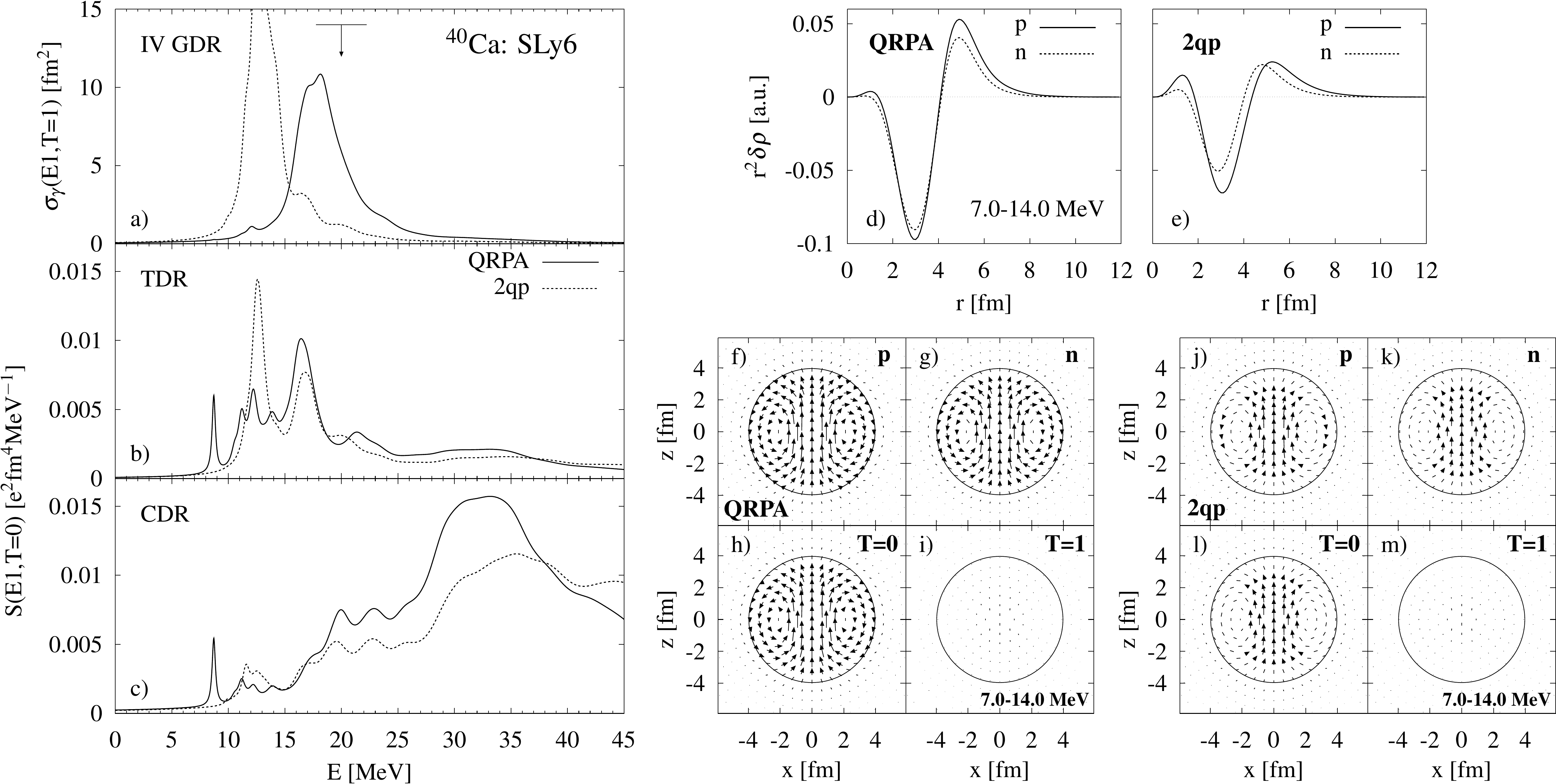}
\caption{(Color online) As in figure 2, but for $^{40}$Ca. Experimental data
\cite{Wy65} for IV GDR
(peak energy and FWHM) are shown.}
\end{center}
\label{fig:40Ca}       
\end{figure*}
\begin{figure*} 
\begin{center}
  \includegraphics[width=1.8\columnwidth,angle=-0]{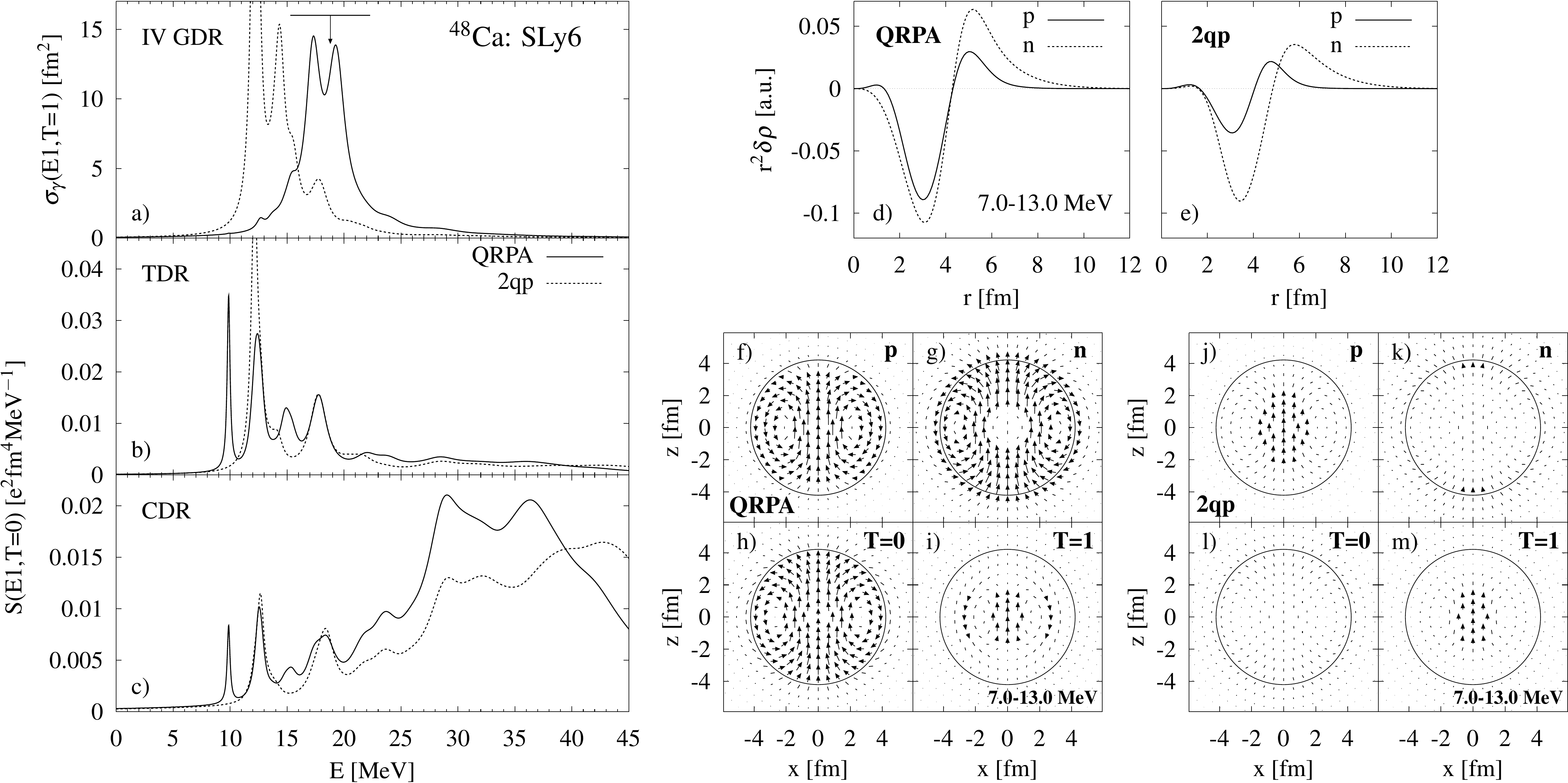}
\caption{As in figure 2, but for $^{48}$Ca. Experimental data
\cite{Kee87} for IV GDR
(peak energy and FWHM) are shown.}
\end{center}
\label{fig:48Ca}       
\end{figure*}
\begin{figure*} 
\begin{center}
  \includegraphics[width=1.8\columnwidth,angle=-0]{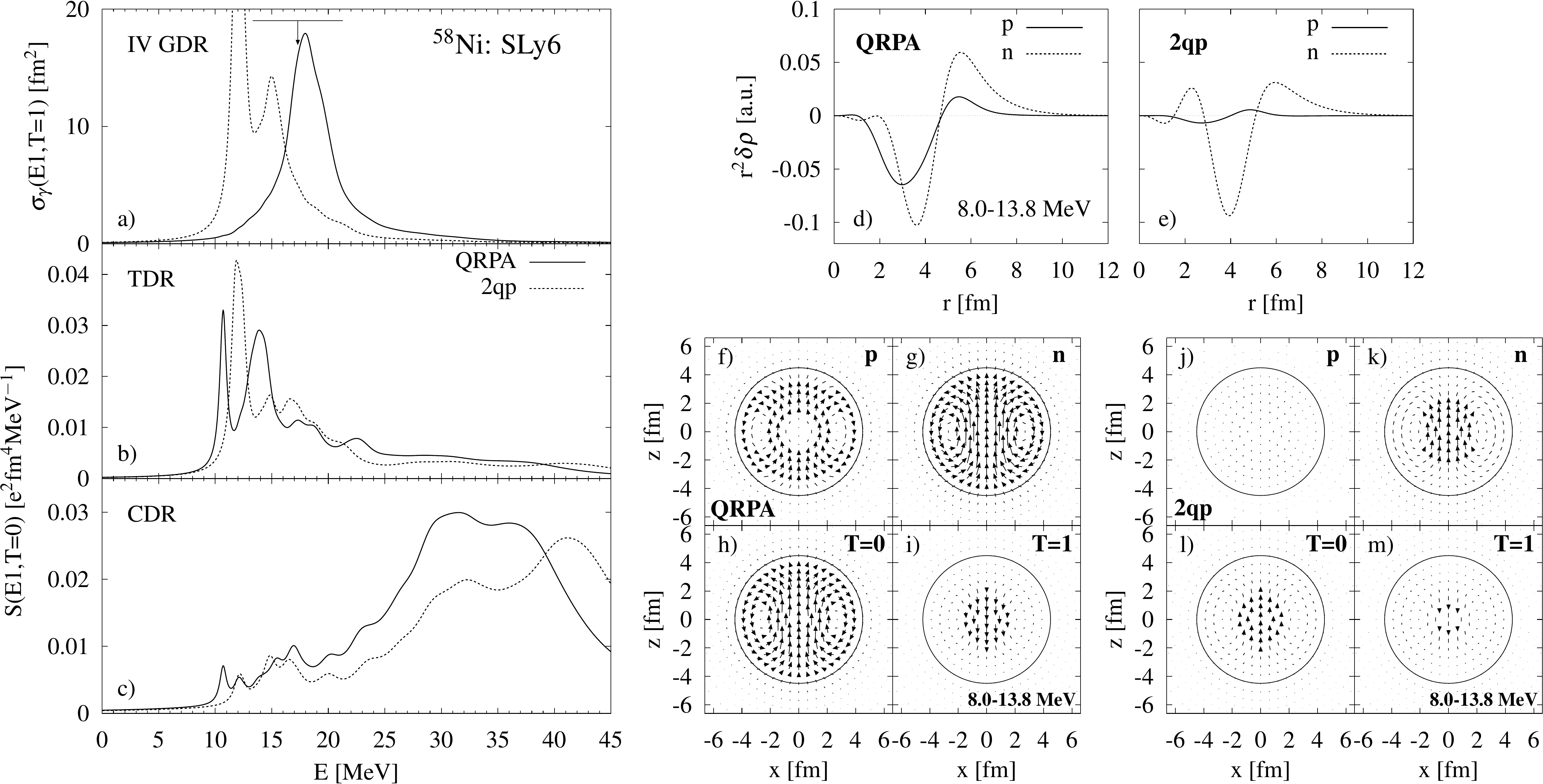}
\caption{As in figure 2, but for $^{58}$Ni. Experimental data
\cite{Fu74} for IV GDR
(peak energy and FWHM) are shown.}
\end{center}
\label{fig:58Ni}       
\end{figure*}
\begin{figure*} 
\begin{center}
  \includegraphics[width=1.8\columnwidth,angle=-0]{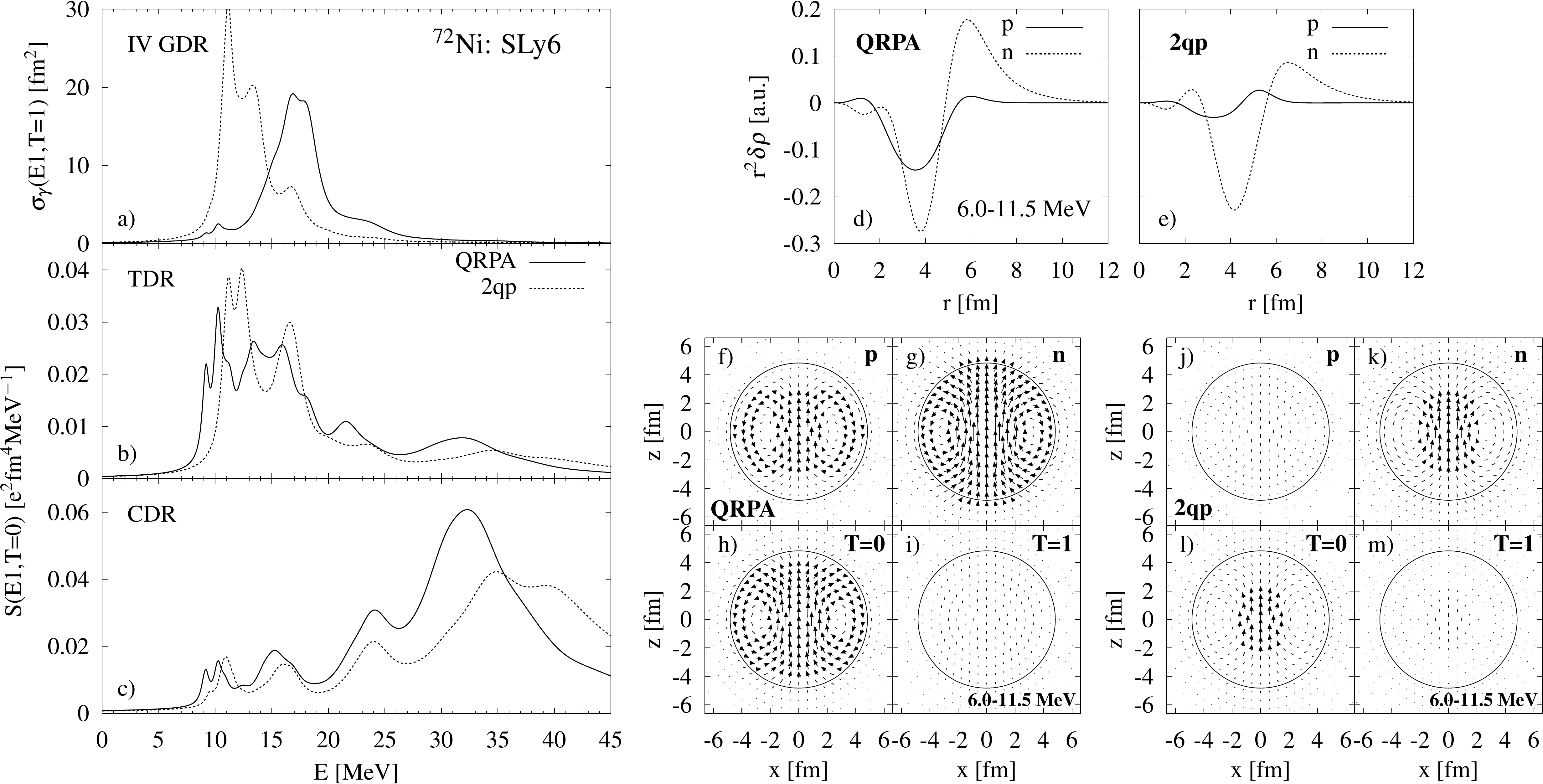}
\caption{As in figure 2, but for $^{72}$Ni.}
\end{center}
\label{fig:72Ni}       
\end{figure*}

Further, we show in Fig. 4 neutron and proton current distributions
together with their angular-dependent divergences
$[\mathrm{div}\; \delta j](r) Y^*_{10}$. Since $Y^*_{10} \sim z/r$, the
divergences in the upper ($z > 0$) and bottom ($z < 0$) parts of
the plots differ only by their sign.
So it is enough to inspect only the $z > 0$ parts of the plots.

Figure 4 shows a strong up-directed flux of protons and neutrons
along z-axis. The flux definitely contains an irrotational flow and
the main divergence spots are mainly  located in its area. It is easy
to see that, for both neutrons and protons, the divergence spots
in Fig. 4 are in one-to-one correspondence with humps in the TD in Fig. 3.
For neutrons, e.g., the divergence spot at $z \sim 1$ fm gives the
positive TD hump at $0.5 < r < 1.5$ fm, the arc-like
spot at $z \sim 4$ fm leads to the negative TD hump at $r= 4-6$ fm, and finally
the arc-like spot at $z \sim 6$ fm results in the TD positive hump
at $r= 6-8$ fm. The impact of the third outer-lying spot is strongly enhanced by
$r^2$-weight, see Fig. \ref{fig:3}(d). Further, in accordance to TD, the proton and neutron
divergences have a different sign at $z \sim 1$ fm and are in phase at
$z \sim 4$ fm.

Note that the first positive hump at 0.5 -- 1.5 fm
is a particular feature of $^{120}$Sn (see for comparison TD
in Figs. 7-14 for other nuclei). Instead, the next two humps, at 4--6 and 6--8 fm,
persist in neutron TD for all considered nuclei.

The appearance of irrotational regions in predominantly toroidal
flow can be explained by simple arguments. As mentioned above,
the exact vortical toroidal current (Hill's  spherical vortex)
${\bf j}_{\rm tor} \sim  \nabla \times \big( \nabla \times
({\bf r}{\rm M}^{\rm com}_{1\mu}({\bf r}))\big)$
is a curl and so cannot produce the irrotational flow.
However, as seen in Fig. 4, the actual toroidal motion
is somewhat squeezed in x-direction.  Indeed the curl centers at
$|x| \sim 3-4$ fm  are rather ovals than circles.  This squeezing leads to
rectification  of the flow in some regions (especially in the central up-flux)
and thus to the appearance of the irrotational motion.
This can also be treated as an admixture of a small irrotational
compressional fraction \cite{Kv11}.
As shown below,
this effect takes place for all considered nuclei.

Finally, the correspondence between neutron TD and TC
demonstrated in Figs. 3-4 allows to conclude that, in nuclei with a
neutron excess, the hump in neutron TD at the nuclear surface
(commonly treated as the proof of the PDR-like collective scheme) can
be naturally produced by the basically toroidal current with some
irrotational fraction.  The PDR scheme generally does not correspond
to the actual TC and is only an oversimplified imitation of the true
nuclear flow.

To check the sensitivity of our results to the choice of the Skyrme
forces, we present in Figs. 5--6 the same variables as in Fig.
\ref{fig:120SnSLy6}, but now for the forces SV-bas and SkM*.
Again we see that our calculations for IV GDR agree with the experimental
data \cite{Lep74}. Almost all the conclusions drawn above for SLy6 for
TDR/PDR come out the same way for SV-bas and SkM*.
The visible difference is that the neutron TC
for SV-bas and SkM* show a clear toroidal flow already in case of mere 2qp
states (see also comparison of SLy6, SV-bas and SkM*  dipole states,
given in Appendix B). This confirms our previous findings \cite{Ne15Dre,Ne18EPJWC}
that the vortical toroidal flow is mainly of single-particle nature.
The $r^2$-weighted TD for SLy6, SV-bas and SkM* look somewhat
different in the 2qp case but acquire the same persistent form in
QRPA, characterized by a negative hump at 4 -- 6 fm and a positive
hump at 6 -- 8 fm. As discussed above, this particular form is
explained by the toroidal distribution of the nuclear
flow in TC.

Since the performance of SLy6, SV-bas and SkM* is similar, we
restrict ourselves in the following to results from SLy6 only.  These
results involve isotopic pairs $^{40,48}$Ca, $^{58,72}$Ni,
$^{90,100}$Zr, and $^{100,132}$Sn. Each pair includes one isotope without
and one isotope with a large neutron excess.

Fig. 7 shows results for $^{40}$Ca.  This light nucleus with N=Z
should not exhibit PDR strength by definition.  Indeed, we see from
panels (d-)(e) that here the proton TD exceeds the neutron one at the
nuclear surface in both 2qp and QRPA cases with the consequence that
dipole strength in the PDR region is not visible.

At the same time, the TD from QRPA has the
same particular form as in $^{120}$Sn. As mentioned above, this form
is explained just by the toroidal flow. Panel (b) shows that, at the
energy region embracing the left wing of IV-GDR (the typical PDR
region), we have, indeed, rather large toroidal strength. The 2qp and
QRPA TC in panels (f)-(m) also show toroidal flow. This flow is
basically isoscalar with similar contributions of proton and neutrons
(which is natural for N=Z nucleus). So TDR does exist in
$^{40}$Ca where PDR strength is absent in principle.
\begin{figure*}  
\begin{center}
  \includegraphics[width=1.8\columnwidth,angle=-0]{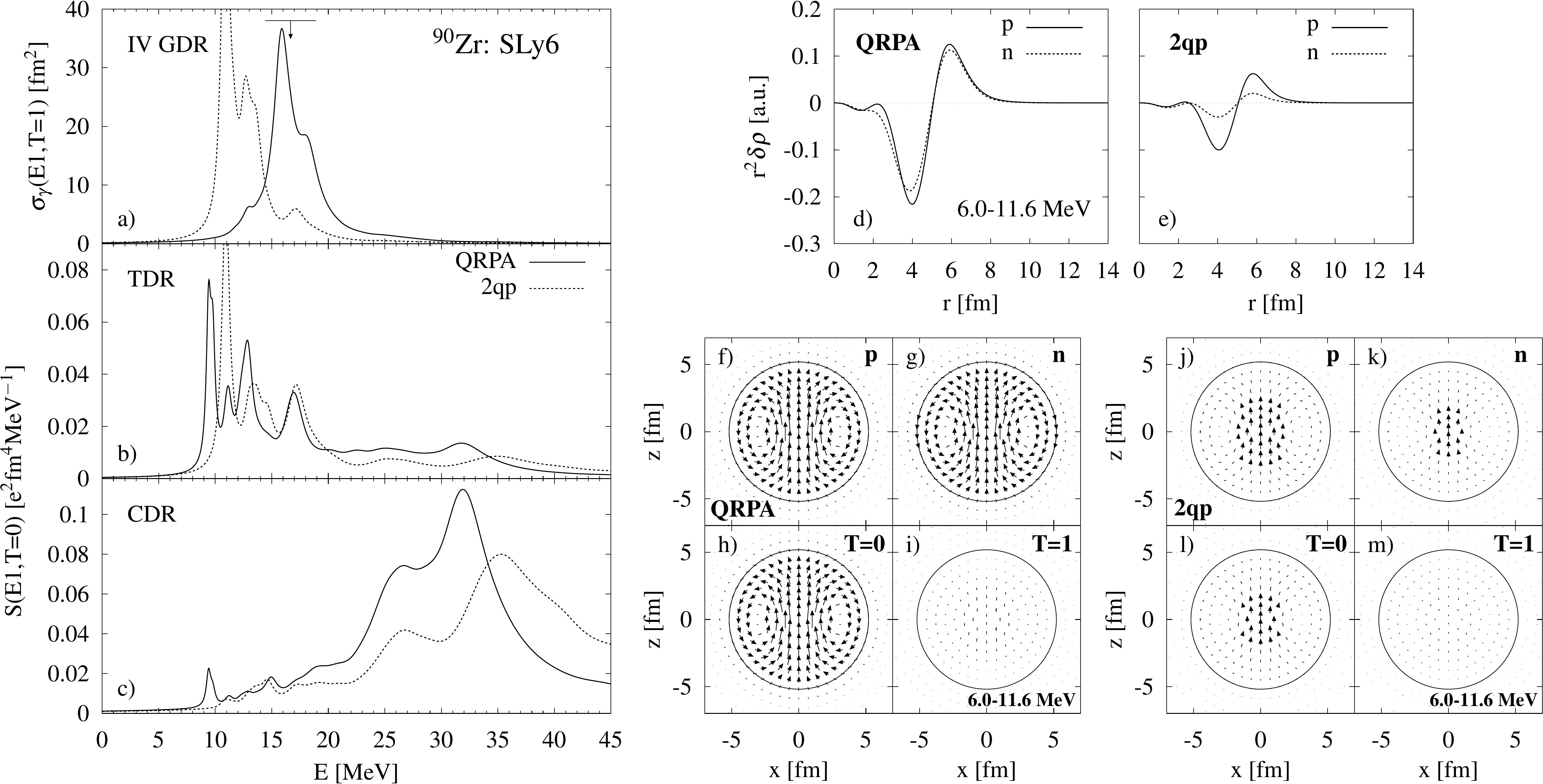}
\caption{As in figure 2, but for $^{90}$Zr. Experimental data
\cite{Ish71} for IV GDR
(peak energy and FWHM) are shown.}
\end{center}
\label{fig:90Zr}       
\end{figure*}
\begin{figure*} 
\begin{center}
  \includegraphics[width=1.8\columnwidth,angle=-0]{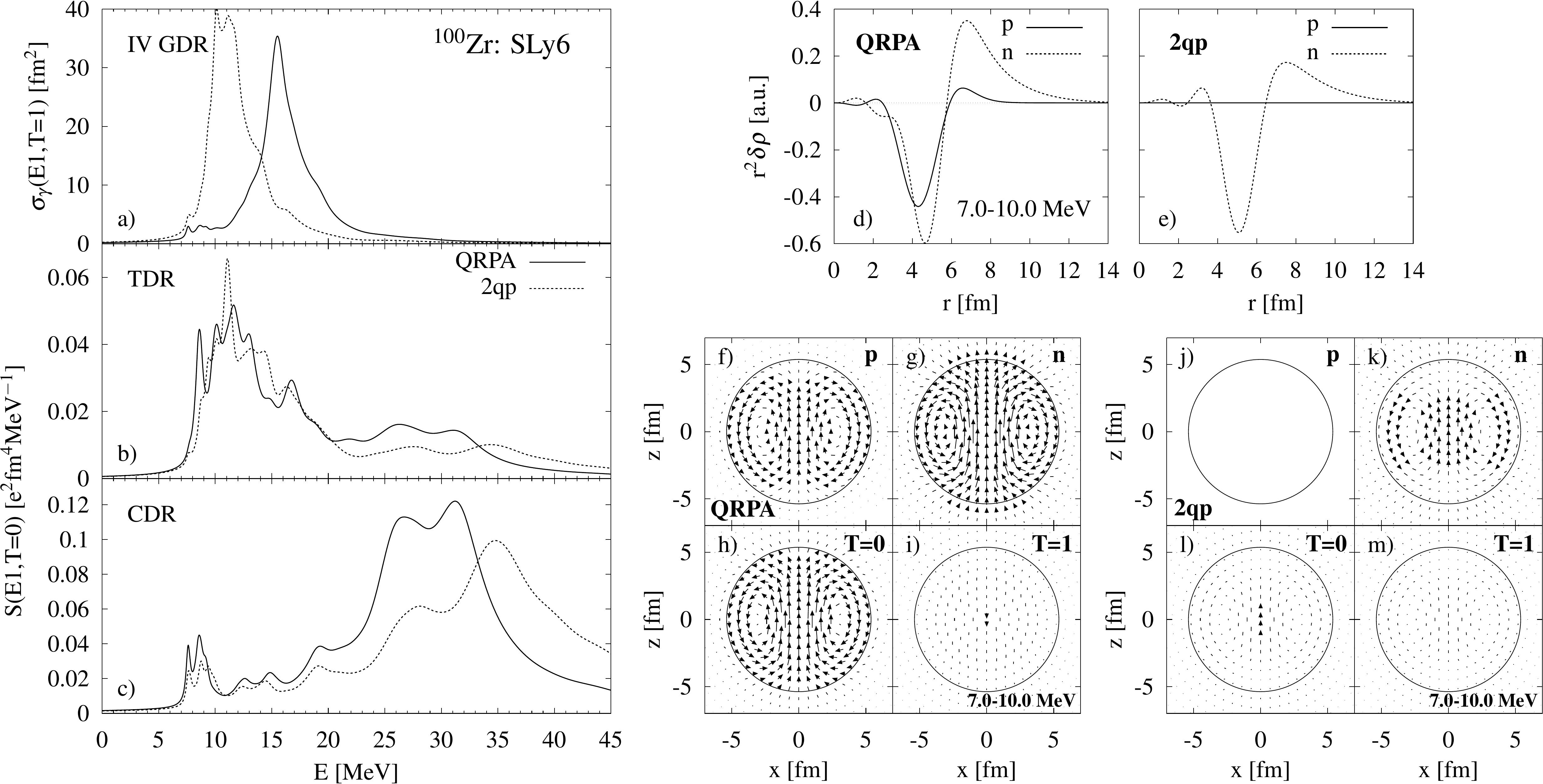}
\caption{As in figure 2, but for $^{100}$Zr.}
\end{center}
\label{fig:100Zr}       
\end{figure*}
\begin{figure*}  
\begin{center}
\includegraphics[width=1.8\columnwidth,angle=-0]{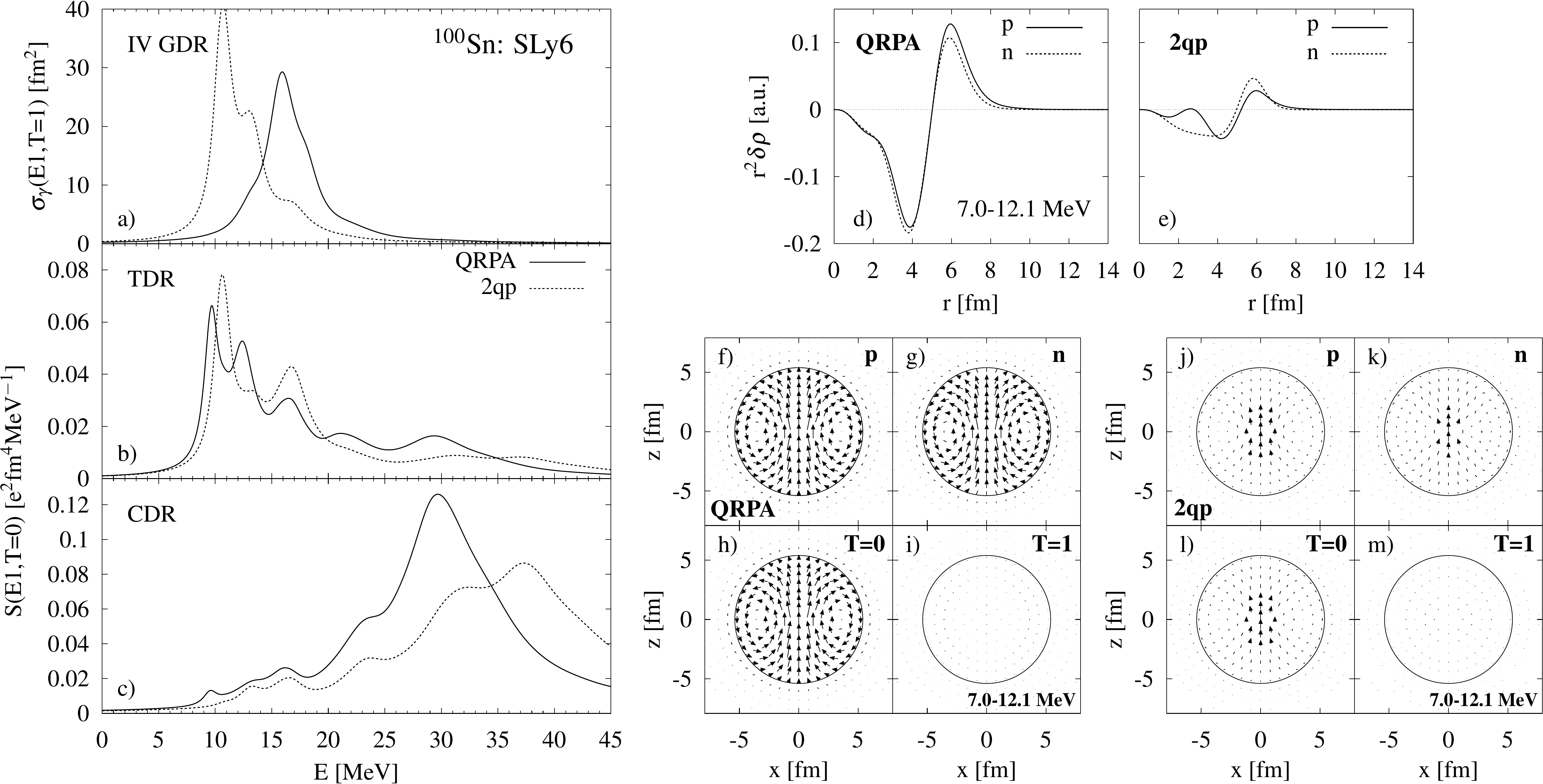}
\caption{As in figure 2, but for $^{100}$Sn.}
\end{center}
\label{fig:100Sn}       
\end{figure*}
\begin{figure*} 
\begin{center}
\includegraphics[width=1.8\columnwidth,angle=-0]{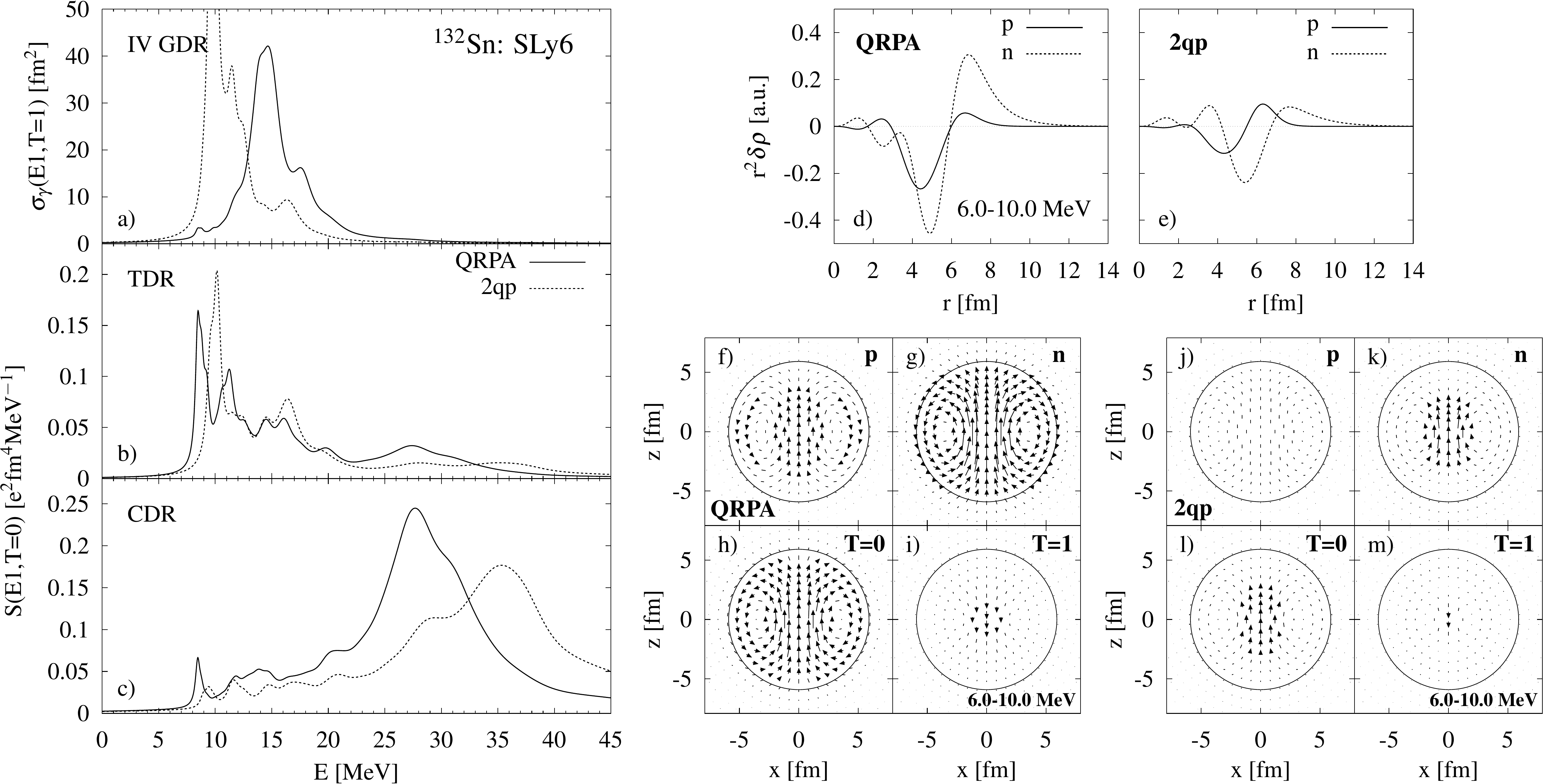}
\caption{As in figure 2, but for $^{132}$Sn.}
\end{center}
\label{fig:132Sn}       
\end{figure*}

Panel (b) shows that the enlarged  TDR strength in $^{40}$Ca
covers a large energy region 8--20 MeV (embracing IV GDR).
This strength is characterized by a few strong peaks. Our analysis shows
that, unlike the lowest vortical IS TDR peak at 10 MeV, the higher peaks have a
more complicated structure. They are formed by dipole states
with different collectivity and
various ratios of vortical/irrotational and IS/IV components.
The vorticity of these states arises  from their major 2qp components.
The vortical fractions contribute to the TDR strength while irrotational
fractions produce E1 strength. The collectivity enhances both strengths.
As seen below, this is a common feature of all the nuclei (with and without
the neutron excess) considered in the present study. Finally note a good
agreement of the calculated IV GDR with experimental data  \cite{Wy65}.

Fig. 8 shows the results for $^{48}$Ca, a nucleus with significant
neutron excess. The picture is basically the same as in $^{40}$Ca. The
only essential difference is the larger contribution of neutrons. This
contribution enhances the toroidal strength at 7-11 MeV (panel (b)),
renders the neutron TD dominant at the nuclear surface (panel (d)), and
essentially enhances the IS toroidal current from QRPA as compared to
the 2qp case. The latter is explained by an enhancement of
the residual interaction due to the neutron excess. Note that we well describe
the experimental data \cite{Kee87} for IV GDR.

The next results shown in Figs. 9-14 for the pairs $^{58,72}$Ni, $^{90,100}$Zr,
and $^{100,132}$Sn confirm (up to detail) all
findings and conclusions drawn above from $^{120}$Sn and $^{40,48}$Ca.
The results for IV GDR are in accordance with the experimental
data for $^{58}$Ni \cite{Fu74} and  $^{90}$Zr \cite{Ish71}.
Finally, our systematics for four isotopic groups with 20$< Z <$50, analyzed
with three Skyrme forces, lead to the unambiguous conclusions that:
\begin{itemize}
\item
   Vortical TDR is a general feature of nuclear dipole excitations.
   It forms a broad structure concentrated
   at 7-20 MeV. Its low-energy IS part sharing the same energy
   with PDR is the main subject of our study.
\item
 The toroidal character of the flow is explicitly confirmed by the TC.
\item
  The toroidal flow is somewhat squeezed, which results in appearance of
    local irrotational regions. These regions contribute to the TD
    and produce an irrotational (e.g. compressional) response,
\item
   TD and toroidal TC have one-to-one correspondence.
\item
   In nuclei with a large neutron excess, the actual flow at the
   nuclear surface can be roughly imitated by the
   picture of an oscillating neutron surface often associated with a PDR.
\end{itemize}

The persistence of the TDR in various nuclei can be explained by simple
arguments. As seen from the panels (a) of Figs. 2 and 5-14, we have a
bunch of unperturbed 2qp dipole strength in each nucleus (dashed
line). This dipole strength is produced by $E1(\Delta {\mathcal N}
=1)$ transitions between the neighbor shells, where ${\mathcal N}$ is
the principle shell quantum number. The energy of the bunch is
typically $E_{\rm sp}\approx 41 A^{-1/3}$ \cite{Har01}. This 2qp strength
has both irrotational and vortical fractions. The IV dipole residual
interaction (corresponding basically to Tassie mode $rY_{1\mu}$) is
strongly repulsive and up-shifts the irrotational strength to form the
large IV GDR but does not affect IS strength, especially its vortical (non-Tassie)
fraction. A large part of the vortical strength remains at $E_{\rm sp}$ and
becomes even dominant there, thus forming the TDR.

Most of the Figs. 2 and 5-14 show a significant enhancement of the
toroidal flow by the QRPA residual interaction. This takes place in
all considered nuclei, except perhaps for $^{40}$Ca and $^{120}$Sn
(SV-bas). Thus we see that, although the toroidal flow is basically of
2qp origin \cite{Kv11,Ne15Dre,Ne18EPJWC,Reinhard_2013,RW87}, it is
considerably enhanced by the residual interaction
which renders the toroidal flow more collective.

Note that the profile of the calculated TD corresponds to toroidal
flow rather than to PDR-like pattern. The main difference lies in the
TD at the nuclear interior.  Following the PDR picture, the nuclear
core with N=Z should move as a whole piece against the cloud of the
excess neutrons at the surface. This should give a basically vanishing TD 
in the whole nuclear interior. However, for all
isotopes considered in our study, the calculated TD exhibit
in the interior a sequence
of positive and negative humps.  As discussed above, such behavior is
typical for dominantly 2qp states. It stays in accordance with the
toroidal distribution of the TC, but contradicts a collective PDR
picture.

\section{Discussion}

For better understanding of the relation between PDR and TDR, it is
instructive to compare their main features.

a) The principle difference between TDR and PDR is that TDR
is mainly vortical while PDR  is irrotational, i.e. their velocities fulfill
the relations $\nabla \cdot \bf{v}_{\rm{TDR}} \approx 0$,
$\nabla \times \bf{v}_{\rm{PDR}}$ = 0. The vortical TDR, unlike the PDR,
does not contribute to the continuity equation and Thomas-Reiche-Kuhn
sum rule.

b) TDR is determined by the convective transition current $\delta \bf{j}_{\rm c}$ while PDR
by the transition density $\delta \rho$. The magnetization part of the nuclear current
can significantly conceal the manifestation of the
convective toroidal excitations in $(e,e')$ and other reactions \cite{Ne_ee}.

c) TDR exists in all the nuclei (with exception of the smallest
ones) independently on whether they have or have not neutron excess
(see arguments at the end of the previous section). Instead, PDR
can exist only in nuclei with a noticeable neutron excess.

d) A genuine PDR assumes a collective flow of the neutron excess.
Instead, TDR has the mean field origin and combines both
unperturbed (2qp) and collective flows.

e) As shown above, the actual PDR strength can be treated as a minor irrotational fraction
of squeezed toroidal dipole states. If so, then PDR fraction can be used as a
doorway state for generation of the toroidal flow in reactions where
vortical excitations are not directly produced.

f) TDR and PDR represent two concepts of dipole motion with
a fixed center of mass. In both modes, the central flow
coexists with the opposite peripheral flow.

g) TDR is a broad structure mainly located at 7--20 MeV. Its low-energy IS part
lies at the PDR energy region. PDR and TDR
are formed from the residues of dipole strength, remaining after the upshift of the main
irrotational dipole strength to the IV giant dipole resonance.

h) Both PDR and TDR have IS and IV components and are, to large extent,
isospin mixed \cite{Kv11,Sa13,Br19}.
In both resonances, IS strength lies lower than IV one. In the present study,
we compare only low-energy IS PDR and TDR parts. Their IV parts
need an independent and, perhaps, more complicated analysis. For example,
as compared to IS TDR, the IV TDR is more affected by the magnetization current
\cite{Kv11}.

Dipole strength in the PDR region is important in many aspects. It is
sensitive to the nuclear
neutron skin and so may help to confine the nuclear symmetry energy
and the isospin-dependent part of the nuclear equation
of state, see e.g. reviews \cite{Pa07,Br19}. Besides, this resonance
is located near the neutron separation threshold $S_n$ and so can have
strong impact on
the astrophysical r-process \cite{Ca10}. A great number of calculations
has been already performed for dipole strength in the PDR region. We
have worked out here the mainly toroidal nature of these
low-energy dipole states, which raises the question whether one would
need to revise all these calculations.
This is not necessary if the calculations employ
the fully detailed 2qp structure of the dipole states (e.g. within
QRPA approaches) which takes properly into account the
balance between major toroidal
and minor irrotational contributions.
In the IV dipole strength, relevant for photo-absorption,
only the minor irrotational fraction
is visible while the toroidal contribution is negligible. In other
reactions (e.g. inelastic electron scattering at large angles),
the effect of the vortical toroidal contribution  can be essential
\cite{Ne_ee,Re_ee,Rich04}. It is thus worthwhile to
inspect the ability of various reactions to identify
the vortical toroidal mode as we will do in the next paragraphs.

The IS TDR constitutes the low-energy part
of IS GDR observed in various isoscalar reactions, first of all in $(\alpha, \alpha')$
(see e.g. \cite{Morsch_80,Adams_86,Davis_97,Clark_01,Youngblood_04,Uchida_PL_03,Uchida_PRC_04}
and reviews \cite{Pa07,Sa13,Br19}). However, our recent Skyrme QRPA calculations
show that this treatment  can be disputed \cite{Rep17EPJA}.  Most probably,
the observed low-energy hump in IS GDR is not a merely TDR structure but a mixture of
of toroidal and compressional fractions.
Moreover, the dipole response
in $(\alpha, \alpha')$ reaction is mainly determined by the transition density
while the toroidal mode is generated by the vortical current which is not
confined by
the continuity equation. So, the $(\alpha, \alpha')$ reaction
is perhaps not suited to search the IS TDR and the direct observation of the toroidal mode
remains an open problem.

An alternative would be to search TDR strength in reactions like
$(^{17}O, ^{17}O')$, $(d, d')$,
high-resolution $(p, p')$ under extreme forward angles,
nuclear fluorescence, etc (see the recent review \cite{Br19}
and references therein). Both polarized and unpolarized projectiles
could be used since the vortical flow, in principle, could cause
a polarization of emerging products. The reactions combining
inelastic scattering with $\gamma$-decay, $(\alpha, \alpha'\gamma),
(^{17}O, ^{17}O'\gamma)$ and
$(p, p'\gamma)$, can be also probed.  A general problem
is that we still do no know definite fingerprints of the vortical flow
in these reactions.
The relevant predictions need yet to be developed within modern theoretical
methods embracing both nuclear structure and reaction mechanisms.

Among promising reactions for search TDR, one should also mention inelastic
electron scattering $(e,e')$  \cite{Ne_ee,Re_ee,Ub71,HB83} or more detailed versions like
$(e,e'\gamma)$ \cite{Pa85,JAP17} and scattering of polarized electrons \cite{Wal04}.
The TDR form factor is transversal and can significantly
contribute to the scattering of electrons at large angles.
However, we meet here a strong competition with
the contribution of the magnetization nuclear current \cite{Ne_ee}. As a result,
we usually have a mixture of the convective (TDR) and magnetization
vortical contributions. Hopefully, the convective TDR form factor can be
extracted from experimental $(e,e')$ data  using a decomposition prescription
\cite{Ne_ee}. The reaction $(e,e'\gamma)$ also looks encouraging since the angle of the
outgoing photon can be sensitive to the vorticity
 of the excited state. However, in reactions with electrons, we also
do not know yet possible fingerprints of the TDR.

In general, it seems that theory and experiment thus far were not able
to propose robust signals for an unambiguous identification
of intrinsic vortical electric dipole modes. This remains still a challenge.

In this connection, the exploration of individual vortical toroidal  $1^-$
states  in light nuclei looks promising. Such states were predicted
in $^{10}$Be, $^{12}$C, $^{16}$O, $^{20}$Ne and  $^{24}$Mg \cite{Ne18PRL,Ne18EPJWC,KE10Be_PRC17,KE12C_PRC18,KE_rew18,KE10Be_arXiv,KE16O_arXiv}.
They can be much easier identified than TDR in heavier nuclei. For example, in $^{24}$Mg,
the individual toroidal state should be the lowest $I^{\pi}K=1^-1$ excitation
near the $\alpha$-particle threshold \cite{Ne18PRL}.

\section{Conclusions}

In this paper, we analyzed from a theoretical perspective the
structure of dipole modes in the low-energy range of 6--13 MeV. The
survey was based on the quasiparticle random-phase approximation
(QRPA) built fully self-consistently on top of a
Skyrme HF+BCS description of the nuclear ground
state \cite{Rep17EPJA,Repcode,Be03,Reinhard_92}.
A large selection of nuclei (Ca, Ni, Zr, Sn) was considered and
three different Skyrme parametrizations were used to probe the generality of the
findings. For each element, we consider at least two isotopes, one
with low (or zero) neutron excess and another one with large neutron
excess in order to probe the impact of excess neutrons predominantly
gathering at the nuclear surface. Particular attention was paid to the
relation between the isoscalar (IS) toroidal dipole resonances (TDR) and
the low-energy
dipole states, often denoted as the IS part of the pygmy dipole
resonance (PDR).  We investigate the structure of the modes in terms
of dipole strength function, transition density (TD), and transition
current densities (TC) visualized as a flow pattern. As further analyzing tool, we
compare the QRPA states with the uncoupled mere two-quasiparticle
(2qp) states which are the BCS generalization of the
one-particle-one-hole  states in  Hartree-Fock. The main piece of analysis is
done for the spherical nucleus $^{120}$Sn.

In all the nuclei, independently of the neutron to proton ratio, we
have found a broad distribution of IS toroidal strength. The
concentration of this strength at 6--20 MeV is denoted as the toroidal
dipole resonance (TDR).  In our study, we address the low-energy part
of the IS TDR located at 6--13 MeV, i.e. at the same energy region as
the PDR.  The persistence of the TDR is explained in terms of shell
structure. The dominant pure 2qp dipole states come from transitions
over one major shell and cover exactly the considered energy range
6-13 MeV. The strong isovector (IV) residual interaction in QRPA
shifts the major fraction of (irrotational) IV dipole strength far up
into the region of the isovector giant dipole resonance (IV GDR) while
the IS vortical states remain much less affected by QRPA and stay in
their original energy range. The IV GDR is thus characterized by
irrotational flow while the dominantly vortical flow remains in the
low energy region. Dipole states with energy above IV GDR can also
exhibit vorticity due to their large 2qp components.

Nuclei with large neutron excess also show some peaks with
considerable IV dipole strength in the PDR region and this comes along
with considerable mixing of IS and IV strength
\cite{End10,Reinhard_2013}.  However, this IV branch of PDR and its
isospin mixing was not scrutinized here and remains an issue for a
subsequent survey. The same holds for the impact of complex
configurations beyond mere 2qp states which may be non-ignorable
\cite{Papa11,Ryezayeva_02,End10} and should be checked.

In all considered nuclei, the flow pattern (from CD) at 6--13 MeV
is obviously toroidal. The toroidal motion is somewhat squeezed, which
creates, in addition to the dominant vortical flow, the local irrotational
regions contributing to the TD.
In the nuclei with neutron excess, the TD has
a neutron hump at the nuclear surface, which is usually considered as
the main justification of a collective picture for the PDR. What we find
is that this hump and other features of TD can be explained by the
toroidal TC.

Altogether, our systematic investigation leads to the unambiguous
conclusion that IS PDR strength is actually an outflow of the
underlying toroidal mode in nuclei with the neutron excess. The
collective pygmy-like picture of an oscillating neutron surface is
only a rough imitation of the actual nuclear flow.
Even after the energy averaging, the calculated TD
exhibit large
fluctuations deep into the inside of the nucleus, which is
characteristic of single-particle structure. 
Toroidal flow is seen already in 2qp current distributions, i.e. has
basically mean-field origin.  The QRPA residual
interaction  enforces  toroidal flow and introduces
some collectivity.



We briefly discussed various reactions, first of all $(\alpha, \alpha')$ and
$(e, e')$,  which could be used to search and identify vortical
toroidal modes.
It is known that in $(e, e')$ reactions the impact of the vortical flow
can be essential if not dominant \cite{Ne18EPJWC,Ne_ee,Re_ee}.
 Unfortunately, theory and experiment have no yet come to
robust proposals for unambiguous identification
of intrinsic vortical electric modes.
Such proposals  should take into account both nuclear structure and
reaction mechanisms and thus may be involved. This remains a
challenge and concentration on exploration of TDR will probably be
the most promising starting point.

\begin{acknowledgement}
The work was partly supported by Votruba-Blokhincev (Czech
Republic-BLTP JINR) and Heisenberg-Landau (Germany - BLTP JINR)
grants. A.R. is grateful for support from Slovak Research and Development
Agency under Contract No. APVV-15-0225, and Scientific Grant Agency
VEGA under Contract No. 2/0129/17. J.K. acknowledges the grant of
Czech Science Agency (Project No. 19-14048S).
\end{acknowledgement}

\appendix
\numberwithin{equation}{section}
\section{Nuclear density and current operators}
\label{Sec:_density_oper}

The density operator reads \cite{Kv11}
\begin{equation}\label{dens_oper}
\hat{ \rho} (\vec r)= \sum_{q =n,p}
e_{\text{eff}}^q \sum_{k \epsilon q} \delta({\vec r} - {\vec r}_k)
\end{equation}
where $e_{\text{eff}}^q$ are proton and neutron
effective charges.

The operator of the full nuclear current consists of
the convective and magnetic (spin) parts \cite{Kv11}
\begin{equation}\label{full_j}
 \hat{\vec j}_{\text{nuc}}(\vec r)=
 \hat{\vec j}_{\text{c}}(\vec r) + \hat{\vec j}_{\text{m}}(\vec r)
= \frac{e\hbar}{m} \sum_{q =n,p}(\hat{\vec j}_c^q(\vec r) + \hat{\vec j}_m^q(\vec r))
\end{equation}
where
\begin{eqnarray}
\hat{\vec j}^q_{\text{c}}(\vec r)&=& -i\frac{e_{\text{eff}}^q}{2}
\sum_{k \epsilon q}(\delta({\vec r} - {\vec r}_k) {\vec \nabla}_k
+ {\vec \nabla}_k \delta({\vec r} - {\vec r}_k)) ,
\\
\hat{\vec j}^q_{\text{m}}(\vec r)&=& \frac{g^q_{s}}{2}  \sum_{k \epsilon q}
{\vec \nabla} \times \hat{\vec s}_{qk} \delta({\vec r} - {\vec r}_k) ,
\end{eqnarray}
and $\hat{\vec s}_q$ is the spin operator, $\mu_N$ is the nuclear magneton,
$g^q_{s}$ is the spin g-factor, $k$ numerates the nucleons.
The isoscalar strength functions (\ref{13})
are calculated with the effective charges $e^{p}_{\text{eff}}=e^{n}_{\text{eff}}=0.5$ and
g-factors  $g^p_s=g^n_s= 0.5 \eta (\bar{g}_{s}^n+\bar{g}_{s}^p)$. Here $\bar{g}_{s}^p$ = 5.58
and $\bar{g}_{s}^p$ = -3.82 are the bare g-factors and $\eta=0.7$ is the quenching parameter
\cite{Har01}.

For TD and TC, we consider four options of the density and current operators (proton, neutron,
isoscalar ($T$=0), isovector ($T$=1)), fixed in
(\ref{eq:averDC}) by the index $\beta$ = p, n, 0, and 1. In TC we use only the convection
current.

Finally, four options are fully determined by the effective charges:
\begin{eqnarray}
\beta &=& p: e^{p}_{\text{eff}}=1, \; e^{n}_{\text{eff}}=0 ,
 \\
\beta &=& n: e^{p}_{\text{eff}}=0 ,\; e^{n}_{\text{eff}}=1 ,
 \\
 \beta &=& 0: e^{p}_{\text{eff}}=e^{n}_{\text{eff}}= 0.5 ,
 \\
\beta &=& 1: e^{p}_{\text{eff}}= \frac{N}{A}, \; e^{n}_{\text{eff}}= -\frac{Z}{A} .
\end{eqnarray}

\section{Low-energy dipole states in $^{120}$Sn}
\label{sec:120Sn}

To demonstrate the accuracy of our numerical results, it is worth
to compare the calculated photoabsorption in $^{120}$Sn with recent experimental
data obtained in inelastic proton scattering under extreme small angles \cite{Kru15_120Sn}
This reaction allows to get the dipole strength below and above the neutron
threshold energy $S_n$ within one experiment. Under almost zero scattering angles,
the reaction is dominated by the Coulomb excitation \cite{PNC19}.
\begin{figure} 
\begin{center}
  \includegraphics[width=0.8\columnwidth,angle=-0]{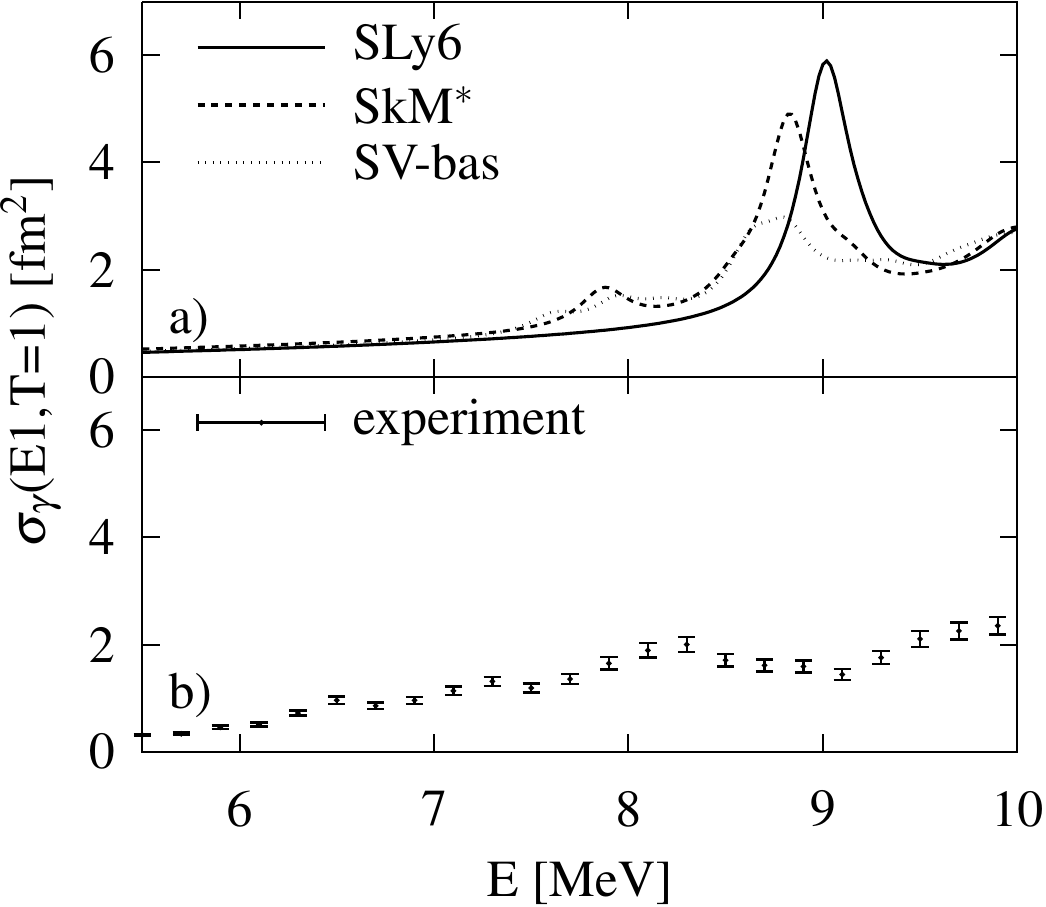}
\vspace{1cm}
\caption{The photoabsorption in $^{120}$Sn at energy  5.5 - 10 MeV:
(a) calculated within QRPA with the Skyrme forces SLy6, SV-bas and SkM*;
(b) extracted from $(p,p')$ data \cite{Kru15_120Sn}.}
\end{center}
\label{fig:3a}       
\end{figure}

In Fig. 15, the photoabsorption calculated with SLy6, SV-bas
and SkM* is compared with the photoabsorption extracted from $(p,p')$ data
 \cite{Kru15_120Sn}. All three Skyrme forces give a significant peak at 8.2-9.5 MeV
which corresponds to the experimentally observed resonance-like structure at 8.3 MeV.
The experimental EWSR strength up to 9 MeV is 2.3(2)$\%$. Our calculations give for the
interval 5.5 - 10 MeV the values 1.3$\%$ (SLy6) and 1.1$\%$ (SV-bas, SkM*).
\begin{table}
\caption{The lowest dipole states with $E <$ 9.2 MeV in $^{120}$Sn, calculated
with Skyrme forces SLy6, SV-bas and SkM*. For each state,
energies $E_{\nu}$, $B(E1)_{\nu}$  values for $0^+0_{\rm gr} \to 1^-_{\nu}0$ 
transitions and maximal 2qp components
(with their contributions to the state norm) are shown.
}
\label{tab-1}       
\begin{tabular}{llll}
\hline
Force &$E_{\nu}$  & $B(E1)_{\nu}$   & main 2qp component \\
  &   MeV & $\rm{e^2 fm^2}$ & \\
\hline
SLy6      & 9.01 & 0.199 & nn $\; [3p_{3/2}\; 3s_{1/2}]$ \; 21$\% $ \\
          &9.19 & 0.015 & nn $\; [3p_{3/2}\; 2d_{3/2}]$ \; 52$\% $  \\
\hline
       & 7.61 & 0.017 & nn $\; [2f_{7/2}\; 1g_{7/2}]$ \; 39$\% $ \\
      & 7.94 & 0.024 & pp $\; [2p_{1/2}\; 3s_{1/2}]$ \; 34$\% $  \\
      & 8.17 & 0.013 & nn $\; [3p_{3/2}\; 2d_{3/2}]$ \; 27$\% $ \\
 SV-bas    & 8.63 & 0.052 & nn $\; [2f_{7/2}\; 1g_{7/2}]$ \; 26$\% $ \\
     & 8.82 & 0.055 & nn $\; [3p_{3/2}\; 2d_{3/2}]$ \; 36$\% $ \\
      & 8.99 & 0.007 & nn $\; [3p_{3/2}\; 3s_{1/2}]$ \; 22$\% $ \\
      & 9.15 & 0.014 & nn $\; [3p_{3/2}\; 2d_{3/2}]$ \; 47$\% $ \\
\hline
       & 7.87 & 0.035 & nn $\; [2f_{7/2}\; 1g_{7/2}]$ \; 75$\% $ \\
     & 8.55 & 0.018 & nn $\; [3p_{3/2}\; 2d_{3/2}]$ \; 57$\% $  \\
 SkM*    & 8.83 & 0.157 & nn $\; [2f_{7/2}\; 1g_{7/2}]$ \; 22$\% $ \\
     & 9.13 & 0.020 & nn $\; [3p_{3/2}\; 2d_{3/2}]$ \; 33$\% $ \\
      & 9.18 & 0.001 & nn $\; [3p_{1/2}\; 3s_{1/2}]$ \; 39$\% $ \\
 \hline
\end{tabular}
\end{table}

In Table I, the features of the calculated dipole states lying below
9.2 MeV are presented. It is seen that for SLy6  dipole states
appear only above 9 MeV. Instead, SV-bas and SkM* dipole spectra start
at 7.5 - 8 MeV and embrace more levels.
The difference can be explained by the effective mass which is much
smaller for SLy6 (0.7) than for  SV-bas (0.9) and SkM* (0.8).
For all three Skyrme forces,
the states are rather collective: contributions of
the largest 2qp components do not exceed 60$\%$ of the state norm
(with exception of 7.87-MeV state in SkM*). This explains the significant effect
of the residual QRPA interaction shown in Sec. 3.
Distributions of E1 strength are different. In SLy6 and SkM*,
$B(E1)_{\nu}$ values are peaked at 9.01-MeV and 8.83-MeV states,
thus  producing the humps visible in Fig. 15 (a).
Instead, in SV-bas the strength  is spread more uniformly.
In general, for SV-bas and SkM*, dipole strengths are more distributed
and less collective than for SLy6. This explains why for SV-bas and SkM*
the toroidal mode is pronounced in TC already in 2qp case.

Though our QRPA calculations roughly reproduce the structure observed
at 8.3 MeV  \cite{Kru15_120Sn}, they overestimate its energy.
Besides, in contrast to the experiment,  the calculations do not give
dipole states below 7 MeV. These discrepancies can be partly explained by the
fact that we do not take into account the coupling with complex configurations (CCC).
Following theoretical estimations in \cite{Kru15_120Sn}, CCC
can be important for the actual distribution of the dipole strength.
At the same time, even models with CCC still  provide quite different
dipole distributions \cite{Kru15_120Sn}.

Table I shows that dipole states in the PDR region significantly vary in
collectivity and structure and so can result in different transition
densities (TD) and transition currents (TC). To overcome this trouble,
TD and TC shown in Sec. 3 are averaged by a special way described in Ref.
\cite{Rep13}. This allows to suppress individual details of the states and
highlight their common general features. Moreover, this partly simulates
the smearing CCC effect.

\section{Acronyms}
\label{sec:acro}

Below we list the acronyms used throughout this paper:
\begin{center}
\begin{tabular}{ll}
 QRPA & quasi-particle random-phase-approximation \\
 GDR  & giant dipole resonance \\
 TDR & toroidal dipole resonance \\
 CDR & compressional dipole resonance \\
 PDR & pygmy dipole resonance (region) \\
 TD & transition density \\
 TC & transition current \\
 IS & isoscalar ($T=0$)\\
 IV & isovector ($T=1$)\\
 E1 & electric dipole \\
 CE & continuity equation
\end{tabular}
\end{center}

\end{document}